\documentclass[twocolumn]{aastex631}
\received{March 9, 2023}
\accepted{\today}
\submitjournal{ApJ}

\shorttitle{Crossing the Rubicon of Reionization with QSOs}
\shortauthors{Grazian et al.}
\graphicspath{{./}{figures/}}

\begin{document}

\title{Crossing the Rubicon of Reionization with $z\sim 5$ QSOs}

\correspondingauthor{Andrea Grazian}
\email{andrea.grazian@inaf.it}

\author[0000-0002-5688-0663]{Andrea Grazian}
\affil{INAF--Osservatorio Astronomico di Padova, 
Vicolo dell'Osservatorio 5, I-35122, Padova, Italy}

\author[0000-0003-4432-5037]{Konstantina Boutsia}
\affil{Las Campanas Observatory, Carnegie Observatories, 
Colina El Pino, Casilla 601, La Serena, Chile}

\author[0000-0003-0734-1273]{Emanuele Giallongo}
\affil{INAF--Osservatorio Astronomico di Roma, Via Frascati 33, I-00078,
Monte Porzio Catone, Italy}

\author[0000-0002-2115-5234]{Stefano Cristiani}
\affil{INAF--Osservatorio Astronomico di Trieste, 
Via G.B. Tiepolo, 11, I-34143, Trieste, Italy}
\affiliation{INFN-National Institute for Nuclear Physics,  
via Valerio 2, I-34127 Trieste}
\affil{IFPU--Institute for Fundamental Physics of the Universe,
via Beirut 2, I-34151, Trieste, Italy}

\author[0000-0003-4744-0188]{Fabio Fontanot}
\affil{INAF--Osservatorio Astronomico di Trieste, 
Via G.B. Tiepolo, 11, I-34143, Trieste, Italy} 
\affil{IFPU--Institute for Fundamental Physics of the Universe,
via Beirut 2, I-34151, Trieste, Italy}

\author[0000-0002-4314-021X]{Manuela Bischetti}
\affil{Dipartimento di Fisica, Sezione di Astronomia,
Universit\`a di Trieste, via G.B. Tiepolo 11, I-34131, Trieste, Italy}
\affil{INAF--Osservatorio Astronomico di Trieste, 
Via G.B. Tiepolo, 11, I-34143, Trieste, Italy} 

\author[0000-0002-0101-6624]{Angela Bongiorno}
\affil{INAF--Osservatorio Astronomico di Roma, Via Frascati 33, I-00078,
Monte Porzio Catone, Italy}

\author[0000-0002-7738-5389]{Giorgio Calderone}
\affil{INAF--Osservatorio Astronomico di Trieste, 
Via G.B. Tiepolo, 11, I-34143, Trieste, Italy}

\author[0000-0002-6830-9093]{Guido Cupani}
\affil{INAF--Osservatorio Astronomico di Trieste, 
Via G.B. Tiepolo, 11, I-34143, Trieste, Italy}
\affil{IFPU--Institute for Fundamental Physics of the Universe,
via Beirut 2, I-34151, Trieste, Italy}

\author[0000-0003-3693-3091]{Valentina D'Odorico}
\affil{INAF--Osservatorio Astronomico di Trieste, 
Via G.B. Tiepolo, 11, I-34143, Trieste, Italy} 
\affil{IFPU--Institute for Fundamental Physics of the Universe,
via Beirut 2, I-34151, Trieste, Italy}
\affil{Scuola Normale Superiore, P.zza dei Cavalieri, 7 I-56126 Pisa, Italy}

\author[0000-0002-4227-6035]{Chiara Feruglio}
\affil{INAF--Osservatorio Astronomico di Trieste, 
Via G.B. Tiepolo, 11, I-34143, Trieste, Italy} 
\affil{IFPU--Institute for Fundamental Physics of the Universe,
via Beirut 2, I-34151, Trieste, Italy}

\author[0000-0002-4031-4157]{Fabrizio Fiore}
\affil{INAF--Osservatorio Astronomico di Trieste, 
Via G.B. Tiepolo, 11, I-34143, Trieste, Italy} 
\affil{IFPU--Institute for Fundamental Physics of the Universe,
via Beirut 2, I-34151, Trieste, Italy}

\author[0000-0003-4740-9762]{Francesco Guarneri}
\affil{INAF--Osservatorio Astronomico di Trieste, 
Via G.B. Tiepolo, 11, I-34143, Trieste, Italy}
\affil{Dipartimento di Fisica, Sezione di Astronomia,
Universit\`a di Trieste, via G.B. Tiepolo 11, I-34131, Trieste, Italy}
\affil{ESO-European Southern Observatory, Karl-Schwarzschild Strasse 2,
D-85748 Garching bei Munchen, Germany}

\author[0009-0004-2597-6146]{Matteo Porru}
\affil{Dipartimento di Fisica, Sezione di Astronomia, Universit\`a di Trieste,
via G.B. Tiepolo 11, I-34131, Trieste, Italy}
\affil{INAF--Osservatorio Astronomico di Trieste, 
Via G.B. Tiepolo, 11, I-34143, Trieste, Italy} 

\author[0000-0003-1174-6978]{Ivano Saccheo}
\affil{Dipartimento di Matematica e Fisica, Universit\`a Roma Tre,
Via della Vasca Navale 84, I-00146, Roma, Italy}
\affil{INAF--Osservatorio Astronomico di Roma, Via Frascati 33, I-00078,
Monte Porzio Catone, Italy}

\begin{abstract}
One of the key open questions in Cosmology is the nature of the sources that
completed the cosmological hydrogen Reionization at $z\sim 5.2$.
High-z primeval galaxies have been long considered the main drivers for
Reionization, with a minor role played by high-z AGN.
However, in order to confirm this
scenario, it is fundamental to measure the photo-ionization rate
produced by active SMBHs close to the epoch of Reionization.
Given the pivotal role played by spectroscopically complete observations
of high-z QSOs, in this paper we present the first results of the RUBICON
(Reionizing the Universe with BrIght COsmological Nuclei)
survey. It consists of a color selected sample of bona-fide $z\sim 5$ QSO
candidates from the Hyper Suprime-Cam Subaru Strategic
Survey. Our QSO candidates have been validated both by photometric
redshifts based on SED fitting and by spectroscopic redshifts,
confirming that they lie at $4.5<z_{spec}<5.2$.
A relatively large space density of QSOs ($\Phi\sim 1.4\times 10^{-8} cMpc^{-3}$)
is thus confirmed at $z\sim 5$ and $M_{1450}\sim -27$, consistent with a pure
density evolution of the AGN luminosity function from z=4 to z=5, with a mild
density evolution rate of 0.25 dex. This indicates that AGN could play a
non-negligible role in the cosmic Reionization.
The Rubicon of Reionization has been crossed.
\end{abstract}

%% Keywords should appear after the \end{abstract} command. 
%% See the online documentation for the full list of available subject
%% keywords and the rules for their use.
%https://astrothesaurus.org/thesaurus/alphabetical-browse/

\section{Introduction}
\label{sec:intro}

During its first billion years, the Universe underwent a major phase
transition for its main baryonic content, the so called Epoch of
Reionization (EoR). The first stars and black holes produced an intense
ultraviolet (UV) radiation that gradually ionized the hydrogen atoms in the
intergalactic medium (IGM), creating ionized bubbles growing for
approximately 1 Gyr, until they fully percolated at $z\le 6$
\citep[e.g.,][]{fan06,fan22,meiksin09,choudhury22}. A very late and
relatively rapid reionization process has been suggested by current
observational constraints \citep[e.g.,][]{planck20}, with a tail-end
extending at $z\sim 5.2$
\citep[e.g.,][]{eilers18,keating20,bosman22,zhu22,jin23}
and lasting for $\Delta z\le 2.8$ \citep{george15,reichardt20}.

The cosmological sources responsible for this disruptive event have
been sought for long time, with conflicting opinions in favor of the
two principal
suspects, i.e. the star-forming galaxies \citep[e.g.,][]{finkelstein19} and the
active galactic nuclei
\citep[e.g.,][]{giallongo12,Giallongo15,giallongo19,boutsia18,grazian18,grazian20,grazian22}. The
mainstream approach to Reionization of all the extragalactic community
has been concentrated for more
than twenty years to demonstrate that the Reionization process started early and
extended in a relatively large redshift interval ($9<z<20$), and that
it is driven only by faint star-forming galaxies \citep[e.g.,][]{lehnert03,bouwens03,bouwens07,dayal18,trebitsch21}.
This was motivated by early results by WMAP \citep{spergel03}.
This scenario, however, has been
shown to be in clear contrast with recent measurements of the ionization status
of the inter-galactic medium (IGM).
The temporal evolution of the neutral hydrogen fraction $x_{HI}$
indicates indeed that it rapidly drops from a value of 1.0 (fully neutral) to
a value of $\sim 10^{-4}$ (almost completely ionized) between $z\sim
7$ and $z\sim 5$ \citep{fan06,hoag19,planck20,fontanot23}. Models assuming
relatively faint star-forming galaxies as the only contributors to the photon
budget of HI ionizing background are thus in tension with the observed
rapid and late reionization process, since these galaxies tend to
start the reionization process too early \citep[e.g.,][]{naidu20}.

One of the first astonishing and transformational results of the James
Webb Space Telescope (JWST) is that the space density of galaxies at
redshift greater than 10 seems to be quite high, possibly similar to or even larger than
the one at redshift 7 \citep[e.g.,][]{labbe22,yan23}, and therefore it is significantly higher than
predictions by numerical simulations \citep[e.g.,][]{haslbauer22}. It is worth
mentioning, however, that these observational results have been based
on the very early data from JWST, that can be affected by (still unknown) calibration issues,
e.g., in data reduction, photometric
calibration, astrometry, as discussed in \citet{finkelstein22,griggio22}.
For example, several galaxies fitted at $z_{phot}>10$ can be instead
extremely dusty starbursts at $z<5$, as shown by \citet{rodighiero23}
with the JWST NIRCam survey in SMACS0723.
A more detailed and mature analysis of the first JWST data is certainly
needed in the future, corroborated also by spectroscopic confirmations
\citep[e.g.,][]{keller22}.
There are two possible explanations to reconcile the large space density of 
galaxies at $z>10$ with a scenario of late reionization:
either most of these galaxies are at $z<10$ or very few ionizing photons are
able to escape from these high-z galaxies into the IGM.

Lastly, a numerous population of faint high-z AGN candidates has been recently
emerging from deep JWST spectra \citep[e.g.,][]{trump22,brinchmann22,wang22,kocevski23,Harikane23,Larson23,Maiolino23},
from compact morphology \citep{ono22} or
from deep MIRI photometry \citep{iani22}. Recent JWST results
from the JADES survey \citep{robertson22} seem to indicate that faint AGN
at $z>4$ are started to be found in deep NIRSpec spectra of normal
star-forming galaxies. For example, \citet{ubler23} and Parlanti et al. (in prep.)
are finding evidences of broad components (FWHM
of $\sim$2000 km/s) for Balmer emission lines (e.g. H$\alpha$, H$\beta$)
for galaxies at $M_{1500}>-22$ at $z>4$, e.g. GDS 273 and GDS 3073 in
the CANDELS GOODS-South field. These powerful outflows can be powered
only by AGN, as shown by \citet{fiore22}.
These two sources have already been shown to host confirmed AGN in the
past \citep{giallongo19,grazian20}, thanks to the X-ray detection by
Chandra for GDS 273 and to the detection of OVI 1032 line in emission for GDS 3073, as
confirmed also by \citet{barchiesi22}. Another results by deep NIRspec
observations is that high-z AGN are quite common at the center
of star-forming galaxies, hinting for a large space density of relatively faint
accreting SMBHs in the primordial Universe, as recently found in the CEERS survey \citep{kocevski23,Harikane23}. This result can have also deep implications for the
early SMBH seeding/collapse scenario, as
recently discussed by \citet{trinca22,trinca23,onoue23,fontanot23}.

The hypothesis of a large space density of faint AGN at $z>4$ and their
possible contribution to the hydrogen reionization is not a new idea.
In the last ten years, indeed, there was a raising consciousness for the
role of AGN as relevant sources of ionizing photons, triggered by
several studies focusing on the faint-end of the AGN luminosity
function (LF) at $z\sim 4-6$. The observations of
\citet{giallongo12,Giallongo15,giallongo19,boutsia18,boutsia21}
and \citet{grazian20,grazian22} have hinted at a larger
than expected number density of faint AGN at $z\ge 4$, which could
imply a significant (if not dominant) contribution of AGN to
the ionizing UV background, if such number densities hold up to higher
redshifts \citep[e.g.,][]{mitra18,fontanot23}. This scenario has been
heavily debated in the past few years, with other studies finding
lower AGN number densities at $z\sim 5-6$
\citep[e.g.,][]{McGreer18,parsa18,akiyama18,kulkarni19,niida20,kimim21,shin22,jiang22}
and recently by \citet{schindler23}.
These contradicting conclusions on the space density of high-z AGN are
likely connected with known problems in the candidate selection, based
either on deep pencil beam surveys (highly affected by strong cosmic
variance effects) or on shallow wide area surveys carried out with
efficient criteria, but typically affected by large incompleteness.

In this paper we exploit the unique combination of deep and wide areal
coverage of the Hyper Suprime-Cam surveys, together with selection
strategy aiming at maximizing the selection completeness, in order to
place a robust measurement of the space density of $L\sim L^*$ QSOs at
$z\sim 5$, close to the main epoch of hydrogen reionization.

The structure of this paper is the following: in Section
\ref{sec:data} we describe the new QSO survey. In Section
\ref{sec:selection} we address the problematic of the selection of
$z\sim 5$ QSOs, their completeness estimate, and the first results of
the follow-up spectroscopic program. In Section \ref{sec:resu} we
derive the QSO luminosity function at $M_{1450}\sim -26.5$, discussing
the evolution of the QSO space density with redshift, and the
derivation of the photo-ionization rate produced by QSOs at $z\sim 5$.
In this section we also check the validity of the recent results
on the AGN Luminosity Function at $z>5$ with a Monte Carlo
simulation. We discuss the reliability of these results in Section
\ref{sec:discussion}, providing the concluding remarks in Section
\ref{sec:conclusion}. Throughout the paper, we adopt H$_{0}$=70 km
s$^{-1}$ Mpc$^{-1}$, $\Omega_{M}$=0.3, and $\Omega_{\Lambda}$=0.7, in
agreement with the $\Lambda$ cold dark matter ($\Lambda$-CDM)
concordance cosmological model. All magnitudes are in the AB
photometric system.

%%%%%%%%%%%%%%%%%%%%%%%%%%%%%%%%%%%%%%%%%%%%%%%%%%%%%%%%%%%%%%%%%%%%%

\section{Data}
\label{sec:data}

\subsection{The RUBICON Survey}

The RUBICON (Reionizing the Universe with BrIght COsmological Nuclei)
survey is an attempt to provide robust constraints to the number
density of $L\sim L^*$ QSOs close to the EoR, i.e. at
$z\sim 5$. The main aim of the RUBICON survey is to measure the
Luminosity Function of $z\sim 5$ QSOs, to derive their contribution
to the photo-ionization rate measured at high-z, and
in preparation for future wide and
deep surveys, e.g. with Euclid, the Roman Space Telescope, and
the Vera Rubin Legacy Survey of Space and Time (LSST).

Due to the dearth of high-z QSOs, especially at bright
magnitudes, large areas of the sky are required in order to build an
efficient and statistically meaningful sample of the rare $z\sim 5$
QSOs. Moreover, at these redshifts even the brightest QSOs
($M_{1450}\le -27$) start to appear as
relatively faint, due to the cosmological dimming effect and to the
strong IGM absorption \citep[e.g.][]{inoue14}. At this aim, we have
adopted as starting database the Third Public Data Release of the
Hyper Suprime-Cam Subaru Strategic Program \citep[HSC-SSP
PDR3,][]{aihara22} in order to search for $z\sim 5$ QSOs at
$M_{1450}\sim -27$. The HSC-SSP PDR3 survey covers an effective area of 34.7
sq. deg. down to magnitudes $\sim$25-27 AB in the grizY bands (Deep
and Ultradeep surveys, hereafter HSC-Udeep), and a larger area of
$\sim$1200 sq. deg. in three extended regions down to a slightly
shallower limits of $\sim$24-26 AB magnitudes (HSC-Wide survey).

The HSC-Udeep survey allows a unique combination of extended (34.7 sq. deg.)
and deep ($G\sim 27$ magnitude depth at 5$\sigma$) multi-wavelength
(grizY) imaging, which is fundamental to overcome the dramatic issues
related to previous surveys (i.e. large cosmic variance effects in
small deep surveys and/or large incompleteness in wide but shallow
surveys). The HSC-Udeep survey is unique at the present time since it
is able to significantly widen the survey discovery space both in
magnitude depth and in the areal coverage. In particular, the
HSC-Udeep survey consists of four separate extragalactic fields (SXDS,
COSMOS, DEEP2-3, ELAIS-N1), which allow us to beat down the cosmic
variance errors to less than 10\%, i.e. a negligible
value\footnote{from Cosmic Variance Calculator \citet{trenti08}
https://www.ph.unimelb.edu.au/$\sim$mtrenti/cvc/CosmicVariance.html} with
respect to the Poissonian noise ($\sim 30-40\%$), that is assumed to
dominate in this survey, given the expected low numbers of high-z QSOs.

In order to complement the HSC-Udeep survey with larger but shallower
areas, with the aim of further reducing the cosmic variance effects,
we select a 108 sq. deg. patch of the HSC-Wide survey, centered around
the SXDS field \citep{furusawa08}. The selected area covers an extended sky
region with Right Ascension in the interval $29.0^{deg}<RA<40.0^{deg}$
and declination in the range $-7.0^{deg}<\delta<+3.0^{deg}$.
The selected
sky patch is slightly larger than the area adopted by \citet{niida20}, allowing
us to make further progresses for the selection of relatively bright
QSOs at $z\sim 5$ and for reducing the cosmic variance effects. The
HSC-Wide survey is $\sim$1 magnitude shallower than the HSC-Udeep
survey. This is not a problem for our purposes, since the QSOs we are
interested on have apparent magnitudes of $\sim$20, and the depths of the
HSC-Wide database are more than adequate for selecting $z\sim 5$ QSO
candidates of $M_{1450}\sim -27$.

%%%%%%%%%%%%%%%%%%%%%%%%%%%%%%%%%%%%%%%%%%%%%%%%%%%%%%%%%%%%%%%%%%%%%

\section{Selection method}
\label{sec:selection}

We have selected from the HSC-Udeep and HSC-Wide surveys all the
sources with magnitudes in the z-band brighter than $magZ=20.0$.
For each source, we have retrieved from the HSC-SSP PDR3 database the PSF
photometry, which is more accurate than Kron magnitudes
for point like objects. At these
apparent magnitudes, indeed, we expect that the contamination from
extended sources is minimal, and the bulk of the sources at $magZ\le
20.0$ are stars or rare QSOs. For the sake of completeness, however,
we have checked that the results in this paper are not affected
whether the Kron magnitudes are used instead of the PSF photometry.

Spectroscopic redshifts for known QSOs in HSC-Udeep and HSC-Wide
surveys have been retrieved from SDSS DR17 \citep{sdssdr17}, after careful
visual inspection of SDSS spectra, from
SIMBAD\footnote{https://simbad.u-strasbg.fr/simbad/sim-fid},
and from NASA/IPAC Extragalactic
Database (NED\footnote{\url https://ned.ipac.caltech.edu/}). In the
HSC-Udeep area we have retrieved from these databases three QSOs
with $4.5<z_{spec}<5.2$,
while in the HSC-Wide region the number of previously known QSOs in this
redshift interval is eight. They are shown in the upper part of Table
\ref{tab:qsoz5udeep} and Table \ref{tab:qsoz5wide}.

\subsection{Selection criteria in the HSC-Udeep survey}

In the HSC-Udeep area there are 84122 sources with $magZ\le
20.0$. This is our starting database for high-z QSOs selection. We
set this relatively bright magnitude limit, $magZ\le 20.0$, both for
HSC-Udeep and HSC-Wide, in order to match our catalog with GAIA DR3
\citep{gaiadr3} and exploit this dataset to reject stars brighter than
$magZ\sim 20$, thanks to the accurate measurements of parallaxes and
proper motions still possible at this flux cut. At fainter
magnitudes, the GAIA DR3 catalog starts to be less complete and going
to fainter magnitude increases the contamination rate expected from
galactic stars. Moreover, we have restricted the analysis to an
apparent AB magnitude of $magZ\le 20.0$, since we want to study the
space density of QSOs close to the luminosity function break ($L\sim
L^*$, i.e. $M_{1450}\sim -27$) and in order to exploit the wide
dynamic range in G-R color uniquely afforded by the deep HSC images.

\begin{figure*}
\includegraphics[width=0.8\linewidth]{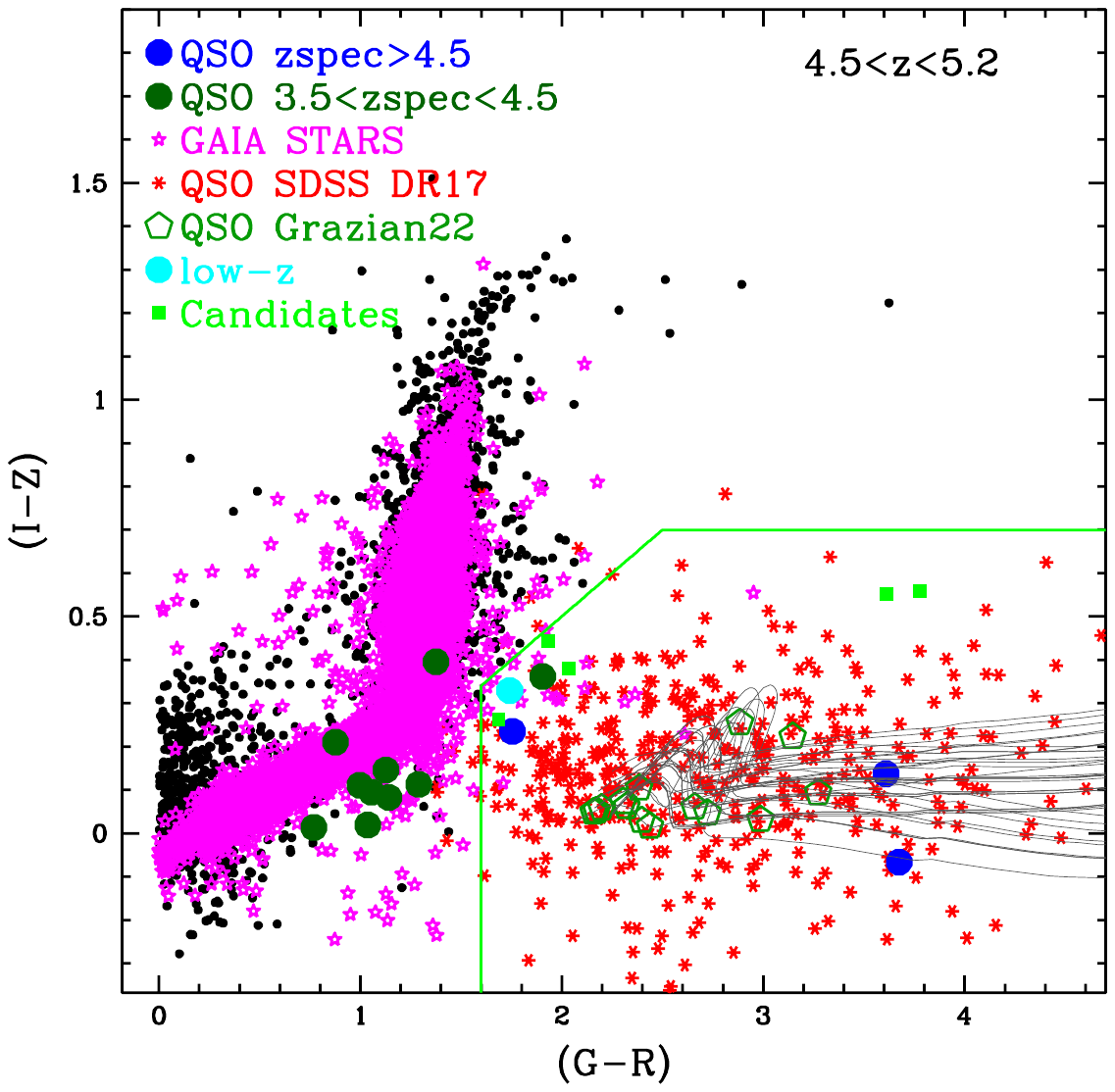}
\caption{The G-R vs I-Z color selection criterion for $z\sim 5$ QSO
candidates in the HSC-Udeep Survey. The black dots are all the objects
with magnitude $magZ\le 20.0$, the red asterisks show the known QSOs with
$4.5\le z_{spec}\le 5.2$ drawn from SDSS DR17 \citep{sdssdr17}. The dark-green
open pentagons are the known very-bright QSOs with $4.5\le z_{spec}\le 5.0$
from the QUBRICS survey \citep{grazian22}, for which synthetic HSC photometry
has been derived starting from the observed SkyMapper photometry through
SED fitting at their fixed spectroscopic redshift. The big blue circles are
the three confirmed QSOs with $4.5\le z_{spec}\le 5.2$ in the HSC-Udeep
area, while the big dark-green circles indicate known QSOs at
$3.5<z_{spec}<4.5$. The cyan circles are confirmed low-redshift objects.
The green lines indicate the color criteria for
selecting bona-fide $z\sim 5$ QSO candidates, while the green squares are
the QSO candidates identified in the HSC-Udeep region. Bona-fide stars
have been identified through accurate measurements of parallaxes and
proper motions
by GAIA DR3 and are marked by magenta starred pentagons. The grey tracks are the
simulated colors of synthetic QSOs in the redshift range $4.5<z<5.2$.}
\label{fig:grizudeep}
\end{figure*}

Bona-fide QSO candidates at redshift $4.5<z<5.2$ have been selected
with the G-R vs I-Z color selection criterion shown in
Fig. \ref{fig:grizudeep}, similar to, but slightly more extended than,
the one previously adopted e.g. by \citet{McGreer18}. The criterion
used here ($G-R\ge 1.6$, $I-Z\le 0.4(G-R)-0.3$, and $I-Z\le 0.75$) is
able to recover $\sim 95\%$ of the SDSS DR17 spectroscopically
confirmed QSOs in the same redshift interval, indicating the
robustness of the adopted color selection. In
Fig. \ref{fig:grizudeep}, the known $z\sim 5$ very bright QSOs,
discovered by \citet{grazian22} in the SkyMapper survey
\citep{wolf18}, sit inside the adopted color selection criteria,
confirming its validity.  The synthetic HSC colors of these QSOs have
been derived through SED fitting to the observed SkyMapper photometry
with AGN libraries fixed at the spectroscopic redshifts of the
objects. The grey tracks in Fig. \ref{fig:grizudeep} are the simulated
colors of synthetic QSOs in the redshift range $4.5<z<5.2$. The mean
IGM absorption by \citet{inoue14} has been adopted for these
tracks. They confirm that the adopted selection criteria is not
affected by strong incompletenesses.

\begin{table*}
\caption{The bright (confirmed and candidate) QSOs at $4.5\le z\le 5.2$
in the HSC-Udeep survey. All these objects have been selected through the
G-R vs I-Z color-color criterion.
}
\label{tab:qsoz5udeep}
\begin{center}
\begin{tabular}{l c c c c c c c c}
\hline
$ID$ & RA & Dec & $magZ_{hsc}$ & z$_{spec}$ & Class & z$_{phot}$ &
Reference z$_{spec}$ & $M_{1450}$ \\
\hline
2480 & 34.685304 & -4.806873 & 19.545 & 4.573 & QSO & 3.78 & SDSS DR17 & -26.63 \\
42780 & 35.302588 & -3.714523 & 19.496 & 5.011 & QSO & 4.98 & SDSS DR17 & -26.89 \\
157404 & 151.161764 & 2.209315 & 19.823 & 5.007 & QSO & 4.97 & \citet{lefevre13} & -26.56 \\
\hline
258 & 244.195589 & 54.324056 & 19.669 & \null & Candidate & 3.78 & \null & \null \\
20375 & 35.334481 & -5.390547 & 18.525 & \null & Candidate & 0.29 & \null & \null \\
131889 & 148.729614 & 1.853699 & 19.940 & \null & Candidate & 3.42 & \null & \null \\
213732 & 244.192594 & 54.342625 & 19.842 & \null & Candidate & 1.49 & \null & \null \\
233735 & 243.709363 & 56.147401 & 19.739 & \null & Candidate & 3.80 & \null & \null \\
\hline
\end{tabular}
\tablecomments{
\\
The QSO $ID=$42780 has been reported also by \citet{McGreer18}.
}
\end{center}
\end{table*}

Among the HSC-Udeep sources brighter than $magZ=20.0$ and falling in
the color-color selection region highlighted in
Fig. \ref{fig:grizudeep}, there are four confirmed QSOs, one with
spectroscopic redshift $z_{spec}<4.5$ (big dark-green circle) and
three with $4.5\le z_{spec}\le 5.2$ (big blue circles).
There are also five QSO candidates with $magZ\le 20.0$ on the
color-color selection region highlighted in Fig. \ref{fig:grizudeep}
(small green squares).

\begin{figure}
\includegraphics[width=\linewidth]{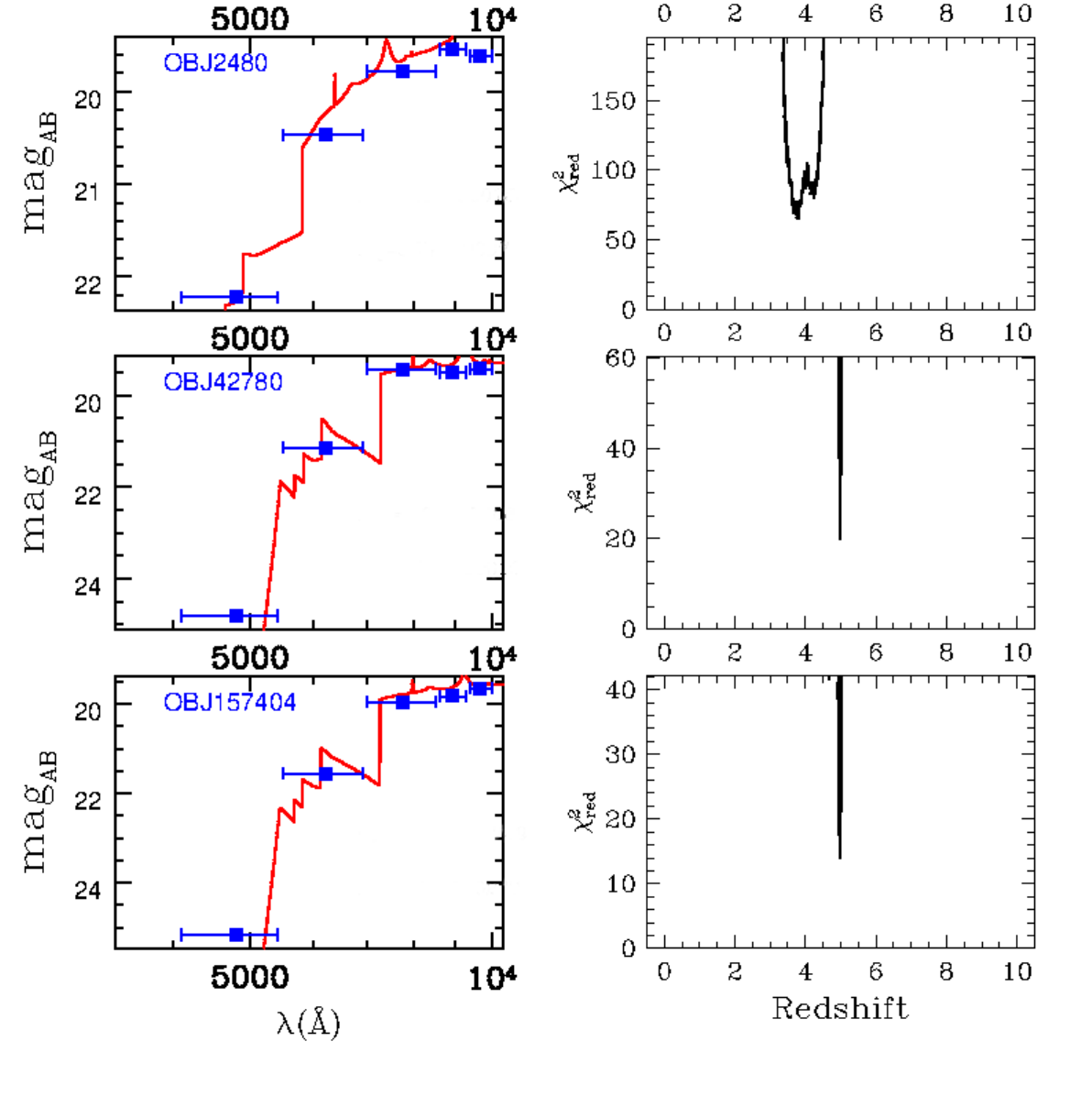}
\caption{The best-fit spectral energy distribution (left) and
the $\chi^2(z_{phot})$ at
different redshifts (right) for three known QSOs at $z_{spec}\sim 5$ in
the HSC-Udeep survey.}
\label{fig:candudeep1}
\end{figure}

For all the known QSOs and the candidates selected through color
criteria, we have computed the photometric redshifts by adopting a
library of synthetic spectra of QSOs taken from LePhare software
\citep{ilbert11}, with the mean IGM opacity of \citet{inoue14}.
These photometric redshifts are based on a $\chi^2$ fitting method
between the observed photometric catalog and the model SED.
The observed photometry, the SED fitting, and the $\chi^2(z_{phot})$
for the three confirmed QSOs at $4.5\le z_{spec}\le 5.2$ in
Table \ref{tab:qsoz5udeep} (top) are shown in Fig. \ref{fig:candudeep1}.
From Fig. \ref{fig:candudeep1}, it is possible to check that these
photometric redshifts give an approximately good indication of the
spectroscopic redshifts for the majority of the known QSOs, but, for
several QSOs, they tend to slightly underestimate the true ones. In
particular, a QSO with $z_{spec}\sim 4.6$ has $z_{phot}\sim 3.8$ in
Table \ref{tab:qsoz5udeep}.
This is mainly due to the fact that the G-R color is not a
strict redshift indicator, but it can also be affected by the variance
of IGM transmission, which is quite large at $z\sim 5$
\citep[e.g.][]{inoue14,worseck14}.
The mismatch between photometric redshifts and spectroscopic ones
could be due to the fact that the former are computed at a mean IGM
absorption \citep[the one by][in this case]{inoue14}, while each
high-z QSO has a stochastic IGM absorption along a given line of sight
which is different from each other, as discussed e.g. in
\citet{cristiani16} and \citet{romano19} for QSOs at $z\sim 4$.

The five QSO candidates in the bottom part of Table \ref{tab:qsoz5udeep}
are shown in Fig. \ref{fig:candudeep2}.
The best-fit SEDs of all these candidates show that the $z_{phot}$ for
these objects are significantly below 4.5, as summarized in Table
\ref{tab:qsoz5udeep}. Moreover, the absence of convincing break in the
observed $G-R$ color for two of these candidates ($ID=$20375 and 213732) could
indicate that they possibly are contaminating stars or low-z galaxies,
as also indicated by their photometric redshifts. The two candidates at
$z_{phot}\sim 3.8$ ($ID=$258 and 233735), instead, could be QSOs at $z_{spec}\sim 4.5$,
as discussed before for the confirmed QSOs, while object $ID=$131889 has
$z_{phot}=3.42$, so it is unlikely that it will turn out to be at $z\ge 4.5$.

Despite the large area covered by the HSC-Udeep survey (34.7 sq. deg.),
the number of confirmed QSOs at $4.5\le z_{spec}\le 5.2$ and $magZ\le
20.0$ is still modest, three, due to the relatively low value of the
QSO space density at high-z and bright magnitudes. Two additional candidates
could be at $z\sim 4.5$, indicating that the current space density of
high-z QSOs provided here is possibly a lower limit.

In order to minimize the cosmic variance
effect, we rely also on a larger portion of the sky, observed by the
HSC-Wide survey near the SXDS field, as we will describe in the
following.

\subsection{Selection criteria in the HSC-Wide survey}

\begin{figure*}
\includegraphics[width=0.8\linewidth]{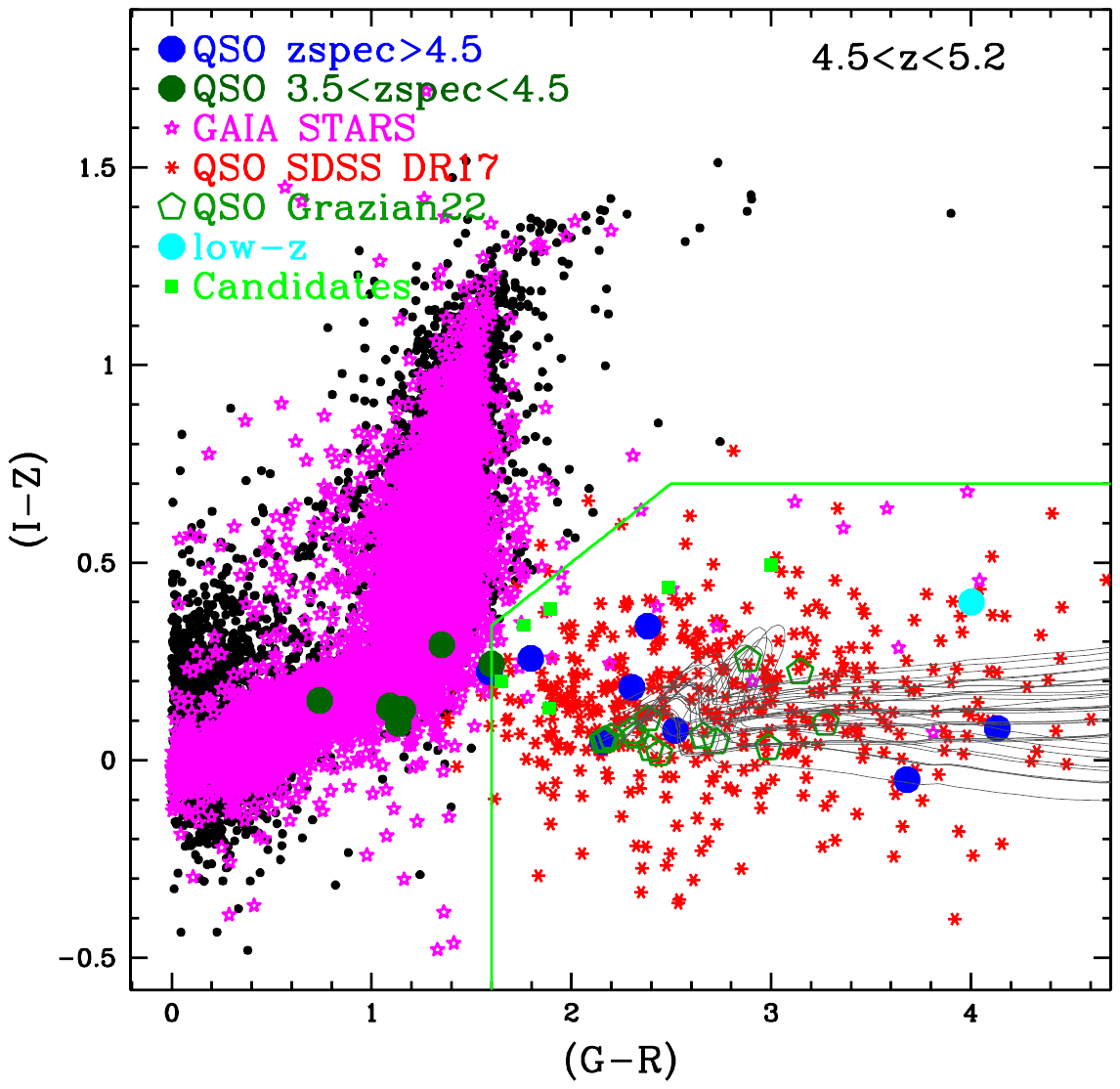}
\caption{The G-R vs I-Z color selection criterion for $z\sim 5$ QSO
candidates in the HSC-Wide Survey. The legend of the symbols are the
same of Fig. \ref{fig:grizudeep}.}
\label{fig:grizwide}
\end{figure*}

In the HSC-Wide region we have selected 245946 sources at $magZ\le
20.0$. This is our starting database for high-z
QSOs selection in the wide area. Fig. \ref{fig:grizwide} shows the
$G-R$ vs $I-Z$ color-color diagram for all sources in the HSC-Wide
area. The black dots are all the objects with magnitude $magZ\le 20.0$,
the red stars show the known QSOs with $4.5\le z_{spec}\le 5.2$
drawn from SDSS DR17. The dark-green open pentagons are the known very
bright QSOs with $4.5\le z_{spec}\le 5.0$ from the QUBRICS survey
\citep{grazian22}. The big blue circles are the known QSOs with
$4.5\le z_{spec}\le 5.2$ in the HSC-Wide area. The green lines
indicate the adopted color criteria for selecting bona-fide $z\sim 5$
QSO candidates, while the green dots are the QSO candidates
identified in the HSC-Wide region.

Eight QSOs with $4.5\le z_{spec}\le 5.2$ have been selected by the
adopted color criteria, while six candidates at $magZ\le 20.0$ still
lack spectroscopic identification, as summarized in Table
\ref{tab:qsoz5wide}. The photometric redshifts for these objects have
been derived as described above for the HSC-Udeep sample. The SEDs and
$\chi^2$ for these sources have been shown in
Fig. \ref{fig:candwide1}, \ref{fig:candwide2}, and
\ref{fig:candwide3}.

\begin{table*}
\caption{The bright QSOs (confirmed and candidates) at $4.5\le z\le 5.2$
in the HSC-Wide survey. All these objects have been selected through the
G-R vs I-Z color-color criterion.}
\label{tab:qsoz5wide}
\begin{center}
\begin{tabular}{l c c c c c c c c}
\hline
$ID_{Wide}$ & RA & Dec & $magZ_{hsc}$ & z$_{spec}$ & Classification & z$_{phot}$ &
Reference z$_{spec}$ & $M_{1450}$\\
\hline
45072$^a$ & 34.685307 & -4.806874 & 19.59 & 4.573 & QSO & 3.71 & SDSS DR17 & -26.59 \\
85212$^b$ & 35.302590 & -3.714526 & 19.45 & 5.011 & QSO & 4.97 & SDSS DR17 & -26.94 \\
50470  & 38.108906 & -5.624829 & 18.55 & 4.565 & QSO & 3.75 & SDSS DR17 & -27.62 \\
155580 & 32.679849 & -0.305117 & 19.12 & 4.732 & QSO & 4.64 & \citet{McGreer13} & -27.13 \\
64891  & 29.079132 & -4.694392 & 19.18 & 4.940 & QSO & 4.79 & \citet{wang16} & -27.17 \\
124850$^c$ & 30.428719 & -1.897370 & 19.56 & 5.021 & QSO & 5.03 & This work & -26.83 \\
157472 & 33.580926 & -1.121369 & 19.92 & 4.628 & QSO & 4.57 & \citet{alam15} & -26.29 \\
220671 & 29.345417 & +2.044333 & 19.31 & 4.503 & QSO & 4.52 & \citet{paris14} & -26.83 \\
\hline
18331  & 34.160449 & -3.627184 & 18.77 & \null & Candidate & 3.90 & \null & \null \\
29348  & 30.086655 & -6.297243 & 19.83 & \null & Candidate & 1.45 & \null & \null \\
124908 & 39.195707 & -2.344237 & 19.75 & \null & Candidate & 3.82 & \null & \null \\
133691 & 37.494808 & -2.247878 & 19.62 & \null & Candidate & 3.52 & \null & \null \\
157609 & 32.563067 & -0.133051 & 19.94 & \null & Candidate & 3.89 & \null & \null \\
188088 & 29.302144 & +0.579305 & 19.81 & \null & Candidate & 3.53 & \null & \null \\
\hline
\end{tabular}
\tablecomments{
The eight objects at the top are confirmed QSOs at $z_{spec}\ge 4.5$, while the
bottom part of the table includes all the $G-R$ vs $I-Z$ color selected QSO candidates.
(a) The confirmed QSO $ID=$45072 is the same QSO with $ID=$2480
in HSC-Udeep survey. (b) The confirmed QSO $ID=$85212 is the same QSO with $ID=$42780
in HSC-Udeep survey. (c) The confirmed QSO $ID=$124850 has been independently discovered
by \citet{yang23}.
\\
}
\end{center}
\end{table*}

Similarly to the case of HSC-Udeep survey, not all of the QSO
candidates in the HSC-Wide area have a photometric redshift above
4.5. One candidate (ID=29348) has a photometric redshift of
$z_{phot}=1.45$ and its SED is a power law without any evident break,
so this is probably a dwarf galaxy at low-z, with
optical colors similar to high-z QSOs. The five remaining
candidates have $z_{phot}\ge 3.5$, indicating that they are potential
QSOs at higher redshifts, given the possible underestimation of the
photometric redshift solutions for QSOs, as we find in the HSC-Udeep
area. A follow-up spectroscopic confirmation of these candidates is
important to derive a complete census of high-z QSOs in this region.

\subsection{A pilot follow-up spectroscopic program in the HSC-Wide area}
\label{sec:spec}

\begin{figure}
\includegraphics[width=\linewidth]{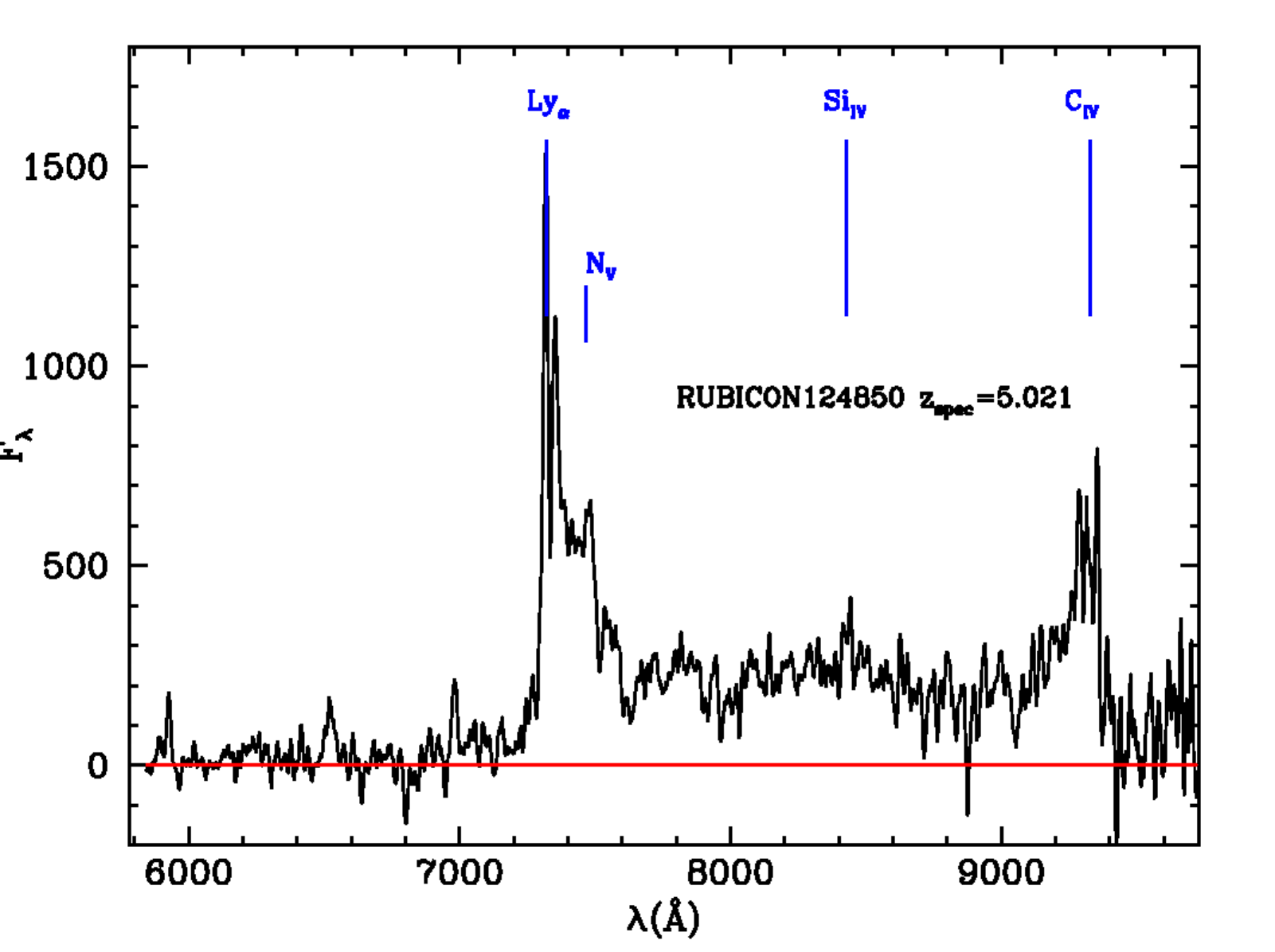}
\caption{The optical spectrum of the newly discovered QSO ID=124850
in the HSC-Wide area.
This object has been confirmed as a QSO at $z_{spec}=5.021$, based on the
identification of broad Lyman-$\alpha$, NV, SiIV, and CIV emission lines.
The red line marks the zero level of the continuum.}
\label{fig:spec124850}
\end{figure}

In the selected G-R vs I-Z region of the HSC-Wide survey, we recover 7
QSOs at $4.5<z_{spec}<5.2$ which were previously known from other
spectroscopic surveys in the literature. Another QSO (ID=124850 at
$z_{spec}=5.021$) has been discovered by a pilot spectroscopic program
carried out by our team at the Magellan telescope with IMACS in
November 2022. The spectrum of this new QSO is shown in
Fig. \ref{fig:spec124850}. This object has been independently
confirmed by recent spectroscopic follow-up of the QSO candidates in
the DESI survey, as described in \citet{yang23}. This new discovery
indicates that the previous spectroscopic surveys in this field have
not yet completed the identification of relatively bright QSO
candidates, and there is still space for improvement, as shown in
Table \ref{tab:qsoz5wide}.

The number of candidates observed within our pilot spectroscopic
project is very limited, and it is not useful at this stage to draw
conclusions on the success rate of our color selection criterion.
This will be carried out in the future, once a larger sky area will be
taken into account. Since the QSO luminosity function has been derived
by considering only the spectroscopically confirmed QSOs, the correct
estimate of the success rate of the survey is not needed for the aims
of the present paper.

%%%%%%%%%%%%%%%%%%%%%%%%%%%%%%%%%%%%%%%%%%%%%%%%%%%%%%%%%%%%%%%%%%%%%

\section{Results}
\label{sec:resu}

\subsection{The Completeness of the RUBICON survey}
\label{sec:compl}

\begin{figure}
\includegraphics[width=0.8\linewidth]{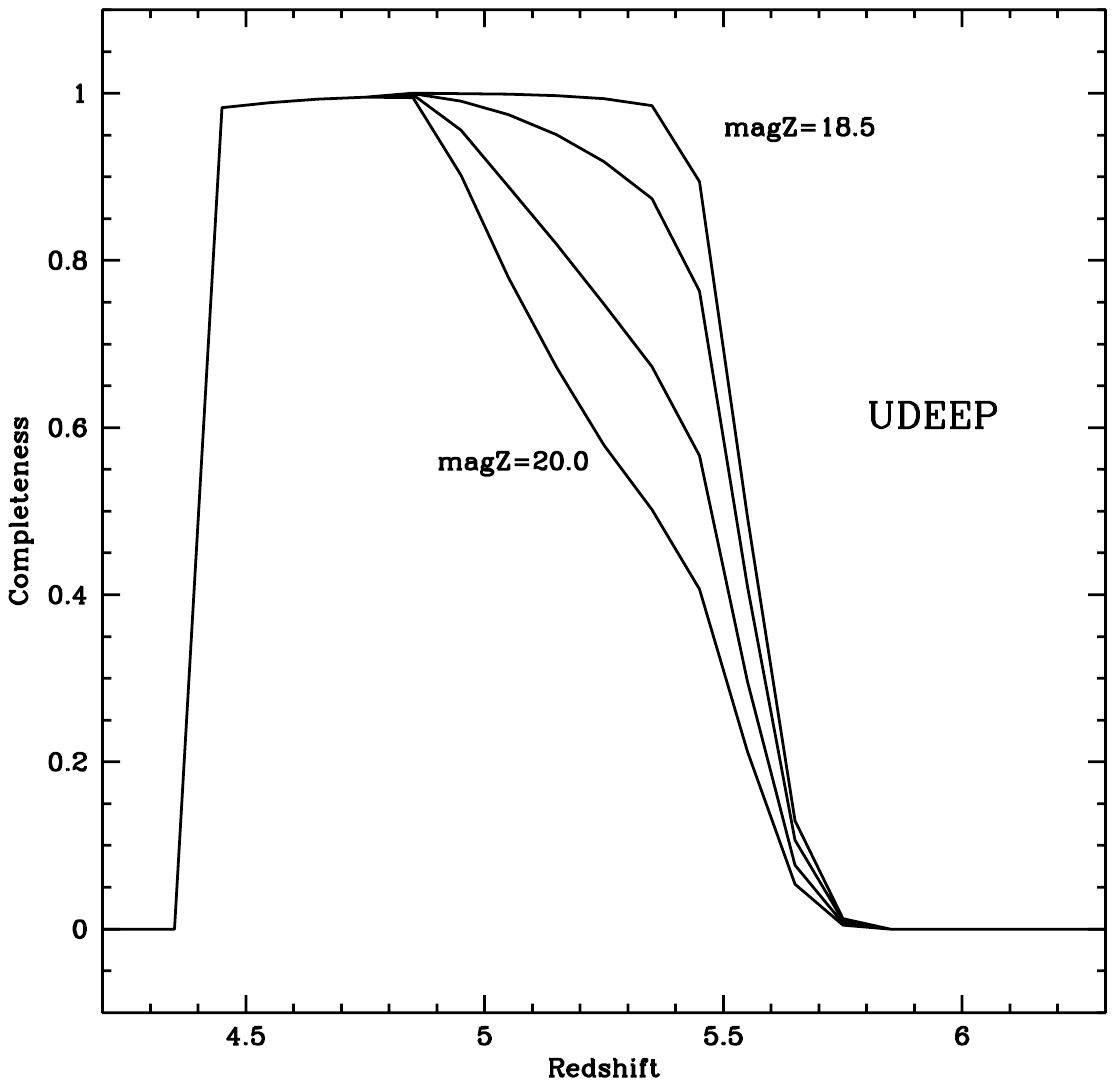}
\includegraphics[width=0.8\linewidth]{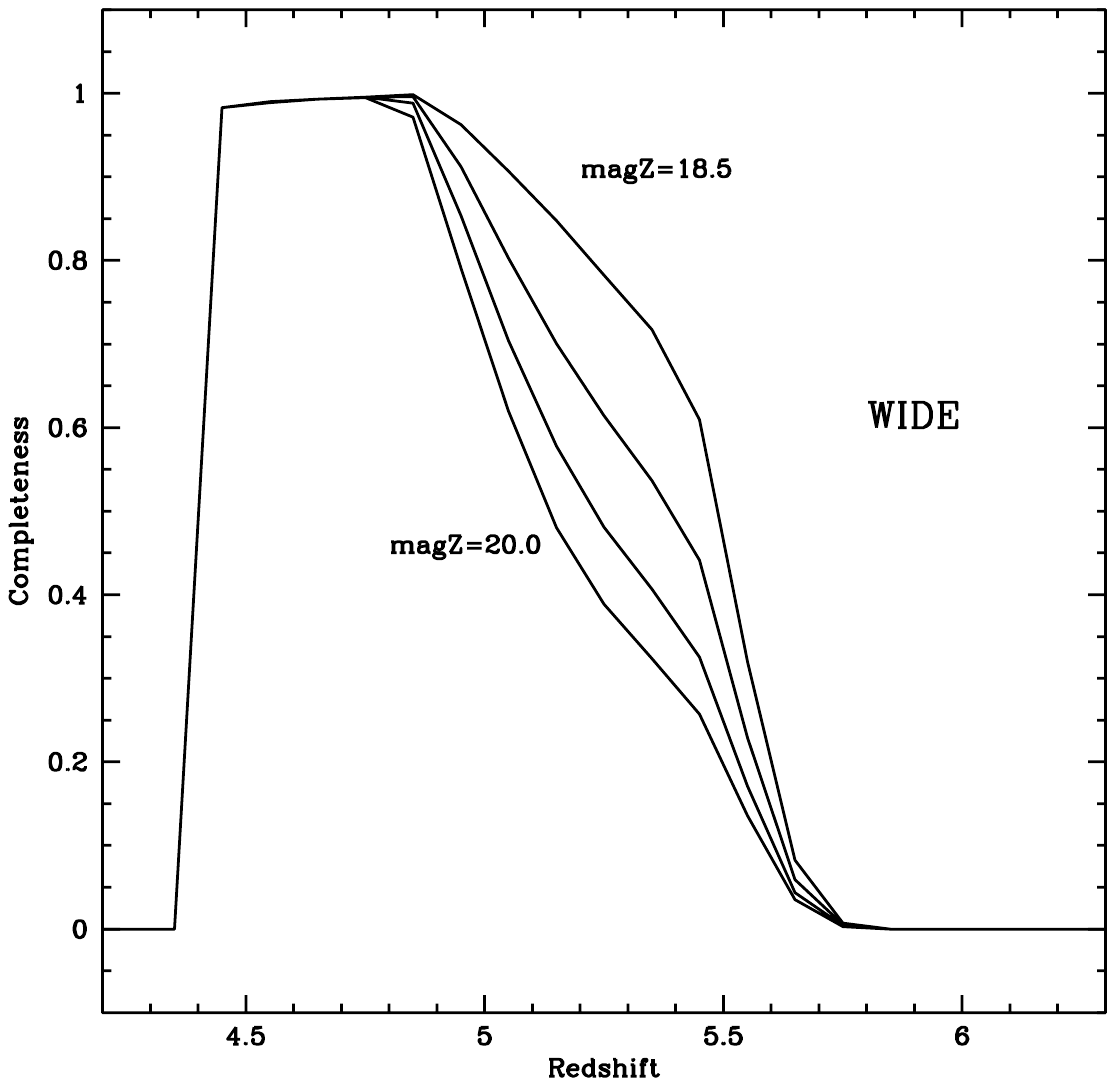}
\caption{The completeness of the HSC-Udeep and HSC-Wide surveys at different z-band magnitudes
(18.5, 19.0, 19.5, and 20.0)
as a function of redshift. The completeness has been derived through simulations, as described in section \ref{sec:compl}.
}
\label{fig:compl}
\end{figure}

In order to provide a fair estimate of the completeness of the RUBICON
survey, we carry out simulations of synthetic QSO colors. We started
from a library of 215 spectra of low-z QSOs from SDSS described in
\citet{fontanot07} and we convolve each of them with IGM transmission
corresponding to 1024 individual line-of-sight simulated using the
\citet{inoue14} formalism, for each redshift bin from $z=4.0$ to
$z=6.1$ with a separation of $\delta z=0.1$. The resulting spectra
have been convolved with the HSC filter curves in grizY to derive the
photometry in the HSC bands. The synthetic magnitudes of these
$\sim$4.8 million QSOs have been perturbed according to the noise
statistics of the HSC images, in order to reproduce the observed
magnitude-error relation for each filter. In particular, we adopt the
magnitude limits at 5$\sigma$ from \citet{aihara22} in order to
reproduce the observed trend of magnitudes versus errors for each
band. We carried out this exercise both for the HSC-Udeep and the
HSC-Wide magnitude limits. We then apply the same color criteria
shown in Fig. \ref{fig:grizudeep} and Fig. \ref{fig:grizwide} to the
synthetic sample affected by noise, and compute the selection function
for different apparent magnitudes in the z-band (from $Zmag=18.5$ to
$Zmag=20.0$) and for different redshifts (from $z=4.5$ to $z=5.2$).
Fig. \ref{fig:compl} shows the completeness of the HSC-Udeep and
HSC-Wide surveys derived through these simulations. It turns out that
the completeness of the adopted color criteria is almost constant,
both in apparent magnitude and in redshift, and the mean value of the
completeness is 94\% and 88\% for the HSC-Udeep and the HSC-Wide
surveys, respectively.

\subsection{The QSO Luminosity Function at $z\sim 5$}

The HSC-Udeep sample of confirmed QSOs at $4.5\le z_{spec}\le 5.2$ in Table
\ref{tab:qsoz5udeep} has been used to compute the luminosity
function of $z\sim 5$ QSOs at $M_{1450}\sim -27$.
The sample in the HSC-Wide area has
been used in order to check the robustness of the
results obtained in the relatively small HSC-Udeep area.
Indeed, the larger area adopted here for the HSC-Wide survey allows us to
decrease the uncertainties associated with the possibly high cosmic variance
in the HSC-Udeep region.

\begin{figure*}
\includegraphics[width=0.8\linewidth]{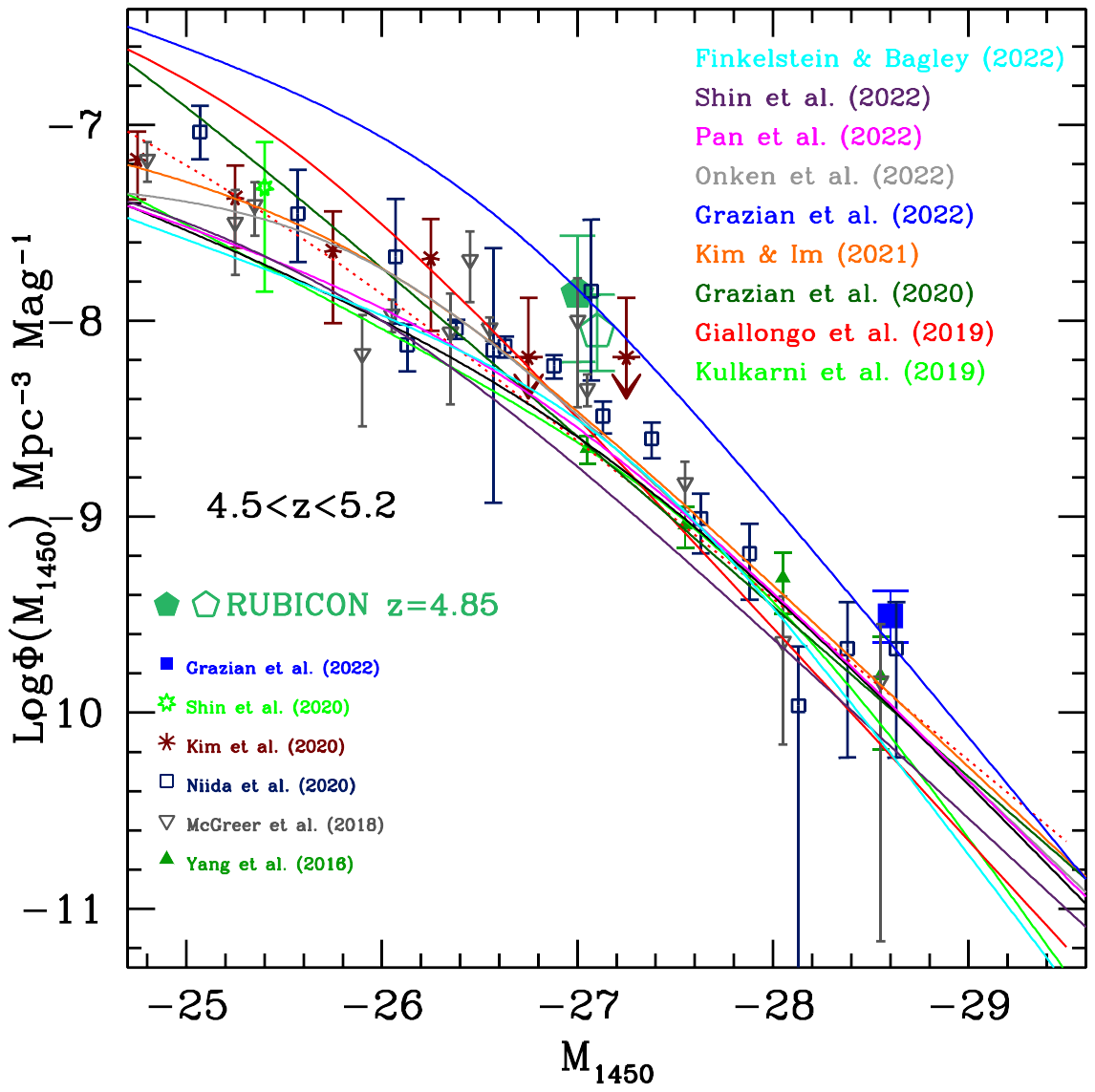}
\caption{The QSO Luminosity Function at $z\sim 5$ from the HSC-Udeep
and HSC-Wide surveys. The filled green pentagon shows the space
density of $z\sim 5$ QSOs in the HSC-Udeep area, while the
open green pentagon shows the space density in the HSC-Wide area, considering only
confirmed QSOs with available spectroscopic redshifts. These
values are a factor of $\sim 2-3$ larger than previous determinations
available in the literature, highlighted by small symbols.
The magenta line is model 4 of \citet{pan22}.
}
\label{fig:lfz5}
\end{figure*}

The absolute magnitudes $M_{1450}$ in Table \ref{tab:qsoz5udeep} and
\ref{tab:qsoz5wide} have been derived for each QSO starting from the
observed magnitudes in the Z band and the spectroscopic redshifts
$z_{spec}$, taking into account the effect of the distant modulus and
the k-correction, as described in \citet{grazian22}. The luminosity
function has been computed as the inverse of the accessible volume,
summed up for all the confirmed QSOs, as described in detail in
\citet{boutsia21} and in \citet{grazian22}. The accessible volume for
each individual QSO has been corrected for the completeness fraction
derived by the simulations described above. In particular, for each
QSO we adopt the completeness correction as a function of the z-band
magnitude and spectroscopic redshift, as shown in
Fig. \ref{fig:compl}. Error bars on the space density of QSOs have
been derived through Poisson statistics, if the number of sources per
bin is above 10, or with the \citet{gehrels86} statistics for smaller
numbers. Fig. \ref{fig:lfz5} shows the QSO Luminosity Function at
$z\sim 5$. The QSO space density obtained in the HSC-Udeep and
HSC-Wide surveys is a factor of $\sim 2$ larger than the one derived
by other surveys in the past
\citep[e.g.,][]{yang16,McGreer18,niida20,kim20,shin20,shin22}. The
blue continuous line shows the best fit of \citet{boutsia21}, evolved
from $z=3.9$ to $z\sim 5$ with a redshift evolution of $\gamma=-0.25$,
as derived by \citet{grazian22}, using the results from the QUBRICS
survey, which is limited to magnitudes $M_{1450}\le -28.3$. Even in
this case, the best fit is a factor of $\sim 2-3$ larger than the
results found by
\citet{kulkarni19,giallongo19,grazian20,kim20,onken22}. The results we
present here for the RUBICON survey are consistent with the
parameterization found by \citet{boutsia21} and the evolution of the
QSO space density derived by \citet{grazian22}, as shown in
Fig. \ref{fig:lfz5}. This result confirms that the evolution of the
QSO luminosity function is consistent with a pure density evolution in
the range $3.5<z<5.5$, with a relatively shallow declining rate ($\sim
-0.25$ dex), as found by \citet{grazian22} and \citet{fontanot23}.

\subsection{Maximum Likelihood simulations}

Maximum Likelihood probabilities have been also computed, adopting as
reference a large collection of publicly available parameterizations
for the QSO Luminosity Function at comparable redshifts. We carried
out a Monte Carlo simulation in order to predict the expected number
of QSOs with $4.5\le z_{spec}\le 5.2$ and magnitude $Z\le 20.0$
observed in the 34.7 sq. deg. of the HSC-Udeep area and in the 108
sq. deg. of the HSC-Wide survey considered here, according to
different parameterizations of the QSO luminosity function proposed at
$z\gtrsim 5$. If the adopted luminosity function is not available exactly at
$z=5.0$, we adopt a pure density evolution with the $dlog\Phi/dz$ parameter quoted
in each paper.
For each best-fit value of a given luminosity function
listed in Table \ref{tab:expectz5}, we randomly extracted the
redshifts and absolute magnitudes $M_{1450}$ for the expected QSOs in
a given area of the sky. We then assign to each simulated QSO a
synthetic SED, randomly taken from the library used to compute the
photometric redshifts described in Section \ref{sec:selection}. The
selected SED is then redshifted to the assigned redshift and
normalized according to the given absolute magnitude $M_{1450}$. This spectrum is
then convolved with the filters adopted for the HSC survey and the
observed magnitudes in the grizY photometric system have been derived
for each simulated QSO. We do not take into account here the effects
of the photometric errors, since at the bright magnitude limit adopted
here ($Z\le 20.0$) the photometric uncertainties should be small. We
then select all the simulated QSOs within the same redshift and luminosity range
of the observed sample, i.e. with $4.5\le z_{spec}\le 5.2$ and
magnitude $Z\le 20.0$. We repeated the simulations for $10^4$
iterations. For each parameterization of the QSO luminosity function
we then derived the mean number $N_{qso}$ of QSOs expected for the
HSC-Udeep and HSC-Wide areas, the standard deviation on $N_{qso}$
for the $10^4$ random realizations and
the probability of recovering at least 9 objects (i.e. the number of observed
QSOs in the HSC-Udeep and HSC-Wide surveys) using the same selection
criteria in these $10^4$ iterations, that is called $Prob(N_{qso}\ge
9)$. Since the spectroscopic confirmation of the candidates
in Table \ref{tab:qsoz5udeep} and \ref{tab:qsoz5wide} is not finished yet,
and given also the incompleteness of the RUBICON survey discussed above,
the observed number of high-z QSOs can be considered as a lower limit
to the real number of $z\ge 4.5$ QSOs. For this reason, we have measured the likelihood
of each luminosity function parameterization by computing $Prob(N_{qso}>=9)$.

First of all, we check the consistency of the luminosity function
of \citet{grazian22} with the observed number of high-z QSOs in the RUBICON survey.
According to Table \ref{tab:expectz5}, the expected number of QSOs (at
$magZ\le 20$ and $4.5\le z_{spec}\le 5.2$) for the HSC-Udeep and
HSC-Wide surveys investigated here should be of $N_{sim}\sim 16$ for
the parameterization of \citet{grazian22}. In the same magnitude and
redshift intervals, we observe 9 bright QSOs. This does not mean,
however, that the \citet{grazian22} prediction is overestimated by a
factor of $\sim 2$, since in these simulations we assume a
completeness of 100\%, while in reality the completeness is of
the order of 88-94\%, as we found in this work. Moreover, the spectroscopic
identification of the QSO candidates in our area is not finished yet,
and other $z>4.5$ QSOs could be discovered among the targets in Table
\ref{tab:qsoz5udeep} or Table \ref{tab:qsoz5wide}. For example, the
candidates $ID=258$ and 233735 in HSC-Udeep and $ID=18331$, 124908, and 157609
have $z_{phot}\sim 3.8$ and all of them could be finally confirmed at $z_{spec}>4.5$,
as we discussed in the previous paragraph.

Table \ref{tab:expectz5} summarizes the expected numbers and the
associated probabilities $Prob(N_{qso}>=9)$ for a number of QSO luminosity functions
appeared recently in the literature. Only the parameterization of
\citet{grazian22} has a probability grater than 80\% to observe 9 or
more QSOs in the HSC-Udeep and HSC-Wide surveys, while all the other
luminosity functions have a probability less than 20\%, due to their
much lower normalization in space density. In particular, the
parameterizations by \citet{pan22} (models 1, 2, 3) have probabilities
$Prob(N_{qso}\ge 9)$ of $\sim 7-20\%$, while all the other luminosity
functions have probabilities less than $\sim 2\%$. We can thus
conclude that all the previous luminosity functions of $z\sim 5$ QSOs
are excluded at 1$\sigma$ confidence level (corresponding to 16\%
probability), while only the QSO luminosity functions of
\citet{grazian22} and model 3 of \citet{pan22} are consistent with the
present data at more than 1$\sigma$. Adopting a more stringent
threshold of 2.3\% confidence level, corresponding to 2$\sigma$, the
following luminosity function are not (statistically) acceptable:
\citet{yang16,McGreer18,kulkarni19,niida20,kim20,kimim21,FB22,onken22,jiang22,shin22,schindler23,Harikane23,Matsuoka23}
and model 4 of \citet{pan22}.

\begin{table*}
\caption{The expected number of QSOs and the maximum likelihood of different QSO luminosity functions for
the HSC-Udeep and HSC-Wide surveys.}
\label{tab:expectz5}
\begin{center}
\begin{tabular}{l r r r}
\hline
Paper & $N_{qso}$ & r.m.s. $N_{qso}$ & $Prob(N_{qso}>=9)$ \\ %& $Prob(|N_{qso}-9|<=3)$ \\
\hline
\citet{grazian22} & 16.02 & 3.99 & 98.03\% \\
\citet{Harikane23} & 0.05 & 0.07 & 0.00\% \\
\citet{Matsuoka23} & 0.06 & 0.25 & 0.00\% \\
\citet{schindler23} & 0.29 & 0.55 & 0.00\% \\
\citet{shin22} w/o K20 & 2.10 & 1.47 & 0.00\% \\
\citet{shin22} w/ K20 & 2.11 & 1.44 & 0.00\% \\
\citet{jiang22} model 75\% c.l. & 0.12 & 0.36 & 0.00\% \\
\citet{jiang22} model 95\% c.l. & 0.14 & 0.39 & 0.00\% \\
\citet{onken22} w/ Kim20 & 3.59 & 1.89 & 1.39\% \\
\citet{onken22} w/ Niida20 & 2.69 & 1.61 & 0.14\% \\
\citet{pan22} model 1 & 4.98 & 2.24 & 7.26\% \\
\citet{pan22} model 2 & 5.49 & 2.30 & 9.83\% \\
\citet{pan22} model 3 & 6.45 & 2.57 & 19.68\% \\
\citet{pan22} model 4 & 2.58 & 1.54 & 0.08\% \\
\citet{FB22} & 3.23 & 1.72 & 0.31\% \\
\citet{kimim21} case 1 & 2.50 & 1.55 & 0.06\% \\
\citet{kimim21} case 2 & 3.01 & 1.73 & 0.42\% \\
\citet{kimim21} case 3 & 3.46 & 1.87 & 0.11\% \\
\citet{kim20} model 1 & 2.71 & 1.65 & 0.28\% \\
\citet{kim20} model 1b & 2.42 & 1.57 & 0.00\% \\
\citet{kim20} model 1c & 2.43 & 1.59 & 0.21\% \\
\citet{kim20} model 2 & 2.89 & 1.73 & 0.14\% \\
\citet{kim20} model 2b & 2.51 & 1.64 & 0.16\% \\
\citet{kim20} model 2c & 3.31 & 1.78 & 0.52\% \\
\citet{niida20} w/ ML fixed $\beta$ & 3.39 & 1.86 & 0.68\% \\
\citet{niida20} w/ $\chi^2$ fixed $\beta$ & 2.97 & 1.74 & 0.28\% \\
\citet{niida20} w/ ML free $\beta$ & 3.53 & 1.82 & 0.74\% \\
\citet{kulkarni19} model 1 & 2.23 & 1.36 & 0.04\% \\
\citet{kulkarni19} model 2 & 1.46 & 1.09 & 0.00\% \\
\citet{kulkarni19} model 3 & 1.33 & 1.00 & 0.00\% \\
\citet{McGreer18} w/ DPL & 1.19 & 1.08 & 0.00\% \\
\citet{McGreer18} w/ paperI & 0.87 & 0.91 & 0.00\% \\
\citet{yang16} model 1 & 0.74 & 0.85 & 0.00\% \\
\citet{yang16} model 2 & 0.75 & 0.86 & 0.00\% \\
\citet{yang16} model 3 & 0.69 & 0.85 & 0.00\% \\
\citet{yang16} model 4 & 0.80 & 0.88 & 0.00\% \\
\hline
\end{tabular}
\tablecomments{Results of $10^4$ simulations.}
\end{center}
\end{table*}

\subsection{The photo-ionization rate produced by $z\sim 5$ AGN}

A recent estimate of the HI photo-ionization rate produced by $z\sim
4$ AGN has been provided by \citet{boutsia21}. The crucial ingredient
in this analysis is the robust determination of the shape and
normalization of the AGN luminosity function in an extended
($-30<M_{1450}<-18$) luminosity range, thanks to the combination of
wide surveys \citep[e.g. QUBRICS,][]{calderone19,boutsia20,guarneri21}
with deep observations
\citep{fontanot07,glikman11,boutsia18,giallongo19}. Based on this
result, \cite{boutsia21} concluded that AGN are able to produce $\sim
100\%$ of the HI ionizing photons at $z\sim 4$, measured through Lyman
forest fitting or via Proximity effect, if the escape fraction of
$\sim$75\% measured for bright QSOs \citep{cristiani16,romano19} is a
common feature also for fainter sources around and below the knee of
the AGN luminosity function \citep{grazian18}.

In our previous work \citep{grazian22}, we find that the space density of
$M_{1450}\sim -28.5$ QSOs at $z\sim 5$ is 3 times higher than previous
determinations, confirming the recently derived results
by \citet{onken22} at
$z\gtrsim 4.5$, by \citet{sch19a,sch19b}, and \citet{boutsia21} at
$z\sim 3-4$. In \citet{grazian22} we also find that the density evolution
of the AGN luminosity function from z=4 to z=5 is milder than
previously reported determinations, e.g. from SDSS survey, and that
the $z\sim 5$ AGN luminosity function shows the same shape of the
$z\sim 4$ one, with a slightly lower normalization, by $\sim 0.25$
dex. Recent results by the CEERS survey seem to confirm this picture,
indicating that the space density of $z\sim 5$ AGN is relatively high
at $M_{1450}\sim -19.5$ \citep{kocevski23,Harikane23}, confirming early
determinations by \citet{giallongo19} and \citet{grazian20}.
In practice, the QSO luminosity function seems to follow
a pure density evolution with a mild rate.

\begin{figure}
\includegraphics[width=\linewidth]{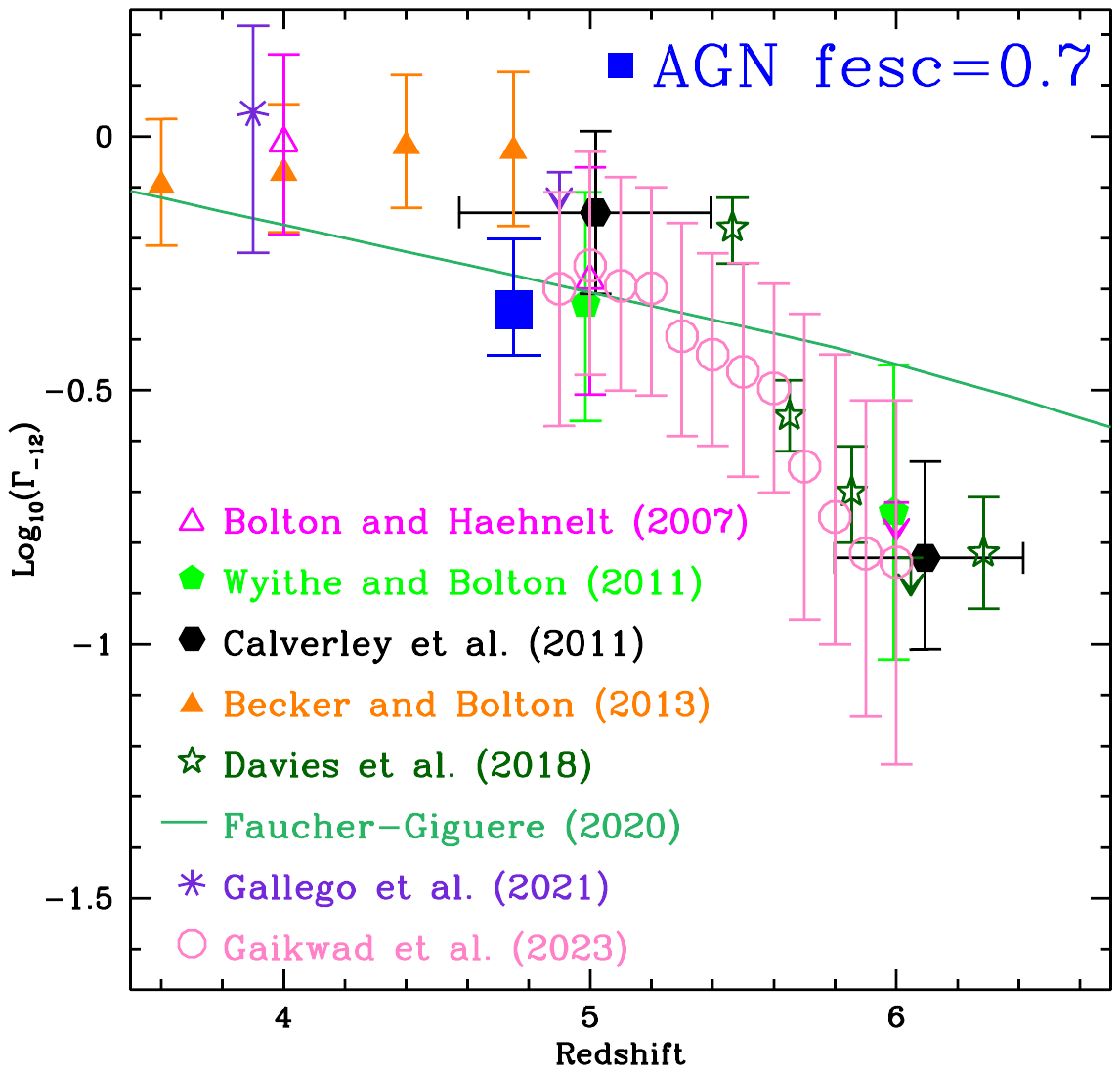}
\caption{The photo-ionization rate of the IGM measured by Lyman forest fitting \citep{Bolton07,WyitheBolton11,BeckerBolton13,davies18,Faucher20,Gallego21,Gaikwad23} and
with the proximity effect \citep{calverley11}. The ionizing background produced by bright QSOs and faint AGN
at $z\sim 5$ is shown by the blue square, assuming a LyC escape fraction of 70\%.
}
\label{fig:gamma}
\end{figure}

The results of this paper confirm the achievements of
\citet{grazian22} at slightly fainter luminosities, $M_{1450}\sim
-27$, close to the knee of the luminosity function, where the space
density of $z\sim 5$ QSOs is still 2-3 times higher than previous
surveys. We assume here that the mean free path of HI ionizing photons
is 17.4 proper Mpc at $z=4.75$ \citep{worseck14} and that the LyC
escape fraction of $z\sim 5$ QSOs is $\gtrsim 70\%$, similar to the
one measured at $z\sim 4$ by
\citet{cristiani16,grazian18,romano19}. It is then possible to compute
the contribution of $z\sim 5$ QSOs to the HI photo-ionization rate, by
integrating the extrapolated luminosity function in the interval
$-30\le M_{1450}\le -18$. Fig. \ref{fig:gamma} shows the photo-ionization rate
produced by $z\sim 5$ QSOs, according to the luminosity function provided in \citet{grazian22},
which is consistent with the data found in this work.
It turns out that bright QSOs and faint AGN
can account for $\sim$50-100\% of the required photon
budget\footnote{This uncertainty is due to the factor of 2 scatter in
the measurement of the hydrogen photo-ionization rate measured through
Lyman forest fitting \citep{Bolton07,WyitheBolton11,BeckerBolton13,davies18,Faucher20,Gallego21,Gaikwad23} or with the proximity effect \citep{calverley11}, as discussed in
\citet{boutsia21} and \citet{grazian22}.} to explain the observed
ionizing background measured close to the end of the EoR, as also
shown by \citet{grazian22} and \citet{fontanot23}.
The photo-ionization rate of AGN derived here,
$\Gamma_{-12}=0.456~s^{-1}$ is consistent with the recent estimate at $z\sim 5$ of \citet{Gaikwad23},
possibly indicating that bright QSOs and faint AGN together are able to provide $\sim 100\%$ of
the ionizing background of the IGM.
This may suggest that QSOs and AGN can play an important role in the cosmological reionization
process of hydrogen.

%%%%%%%%%%%%%%%%%%%%%%%%%%%%%%%%%%%%%%%%%%%%%%%%%%%%%%%%%%%%%%%%%%%%%

\section{Discussion}
\label{sec:discussion}

\subsection{Comparison with recent surveys of $z\sim 5$ QSOs}

It is worth asking why recent surveys of $z\sim 5$ bright QSOs,
e.g. \citet{niida20} or \citet{shin22}, found a QSO number density which
is significantly lower than our values. In order to carry out a fair
comparison, we carefully checked the criteria adopted by these two
surveys.

Starting from \citet{niida20}, we have checked that our QSO ID=157404 in
the HSC-Udeep survey (see Table \ref{tab:qsoz5udeep}) is outside their
area. The other two QSOs at $z>4.5$ in our HSC-Udeep survey, instead, are
part of the area covered by \citet{niida20}. The first object,
ID=2480, does not satisfy their $R-I$ color criterion,
i.e. $R-I>1.0$. Object ID=42780 in Table \ref{tab:qsoz5udeep}
satisfies all their color criteria and it is plausibly included in the
\citet{niida20} QSO sample (their parent QSO sample is not publicly available).
From this comparison, we can conclude that
the QSO luminosity function of \citet{niida20} could be 50\%
incomplete at $M_{1450}\sim -27$, and, by applying this correction factor,
their QSO space density turns out to be compatible with our results in
Fig. \ref{fig:lfz5}.

In order to compare our results with \citet{shin22}, we download the full
HSC-Udeep
survey from the HSC PDR2 repository. QSO ID=157404 in Table
\ref{tab:qsoz5udeep} is outside their area, while the other two $z\ge 4.5$
QSOs, ID=2480 and ID=42780, have been recovered by the PDR2 release. It is
difficult to reproduce the selection criteria of \citet{shin22}, but
it is clear that these two objects are missing from their final sample. We
can thus conclude that their incompleteness could be substantial at $M_{1450}\sim
-27$. It is thus not surprising at all that they find a much lower QSO
number density than our results in Fig. \ref{fig:lfz5}.

The reason why previous surveys found a low number density of $z>4.5$ QSOs
could be due to their incompleteness.
Out of the 8 known QSOs at $z>4.5$ in the HSC-Wide survey (Table \ref{tab:qsoz5wide}),
only 3 have been discovered by SDSS,
while the other QSOs have been found by different surveys. Thus, the total number of $z>4.5$ QSOs is
large in the considered area, but the individual surveys in the past
only recovered a small fraction of them, possibly due to their efficient but strict selection
criteria, which lost a non-negligible fraction of high-z objects.
As an example, if we had adopted only a sub-sample of the color criteria by \citet{McGreer18}, i.e. $G-R\ge1.8$ and $I-Z\le0.5$,
we would have selected 2 out of 3 QSOs at $z_{spec}\ge 4.5$ in the HSC-Udeep survey and 6 out of 8
QSOs at $z_{spec}\ge 4.5$ in the HSC-Wide area, with a further reduction of 66-75\% of the survey completeness. If we apply all the color criteria of \citet{McGreer18}, i.e. $G-R\ge1.8$, $I-Z\le0.5$, $R-I\ge 1.3$, $I-Z\le 0.15+0.875*(R-I-1.30)$, we would have selected 2 out of 3 QSOs at $z_{spec}\ge 4.5$ in the HSC-Udeep survey and only 3 out of 8
QSOs at $z_{spec}\ge 4.5$ in the HSC-Wide area. In this case, by applying the more stringent color criteria by \citet{McGreer18}, the completeness drops down to a level of 38-66\%.

As a general comment, surveys searching for high-z QSOs are usually
very efficient, above 50\% \citep[e.g.,][]{shin22}, but their
completeness level could be very low, as already pointed out in
\citet{grazian20,boutsia21} and in this paper. This incompleteness is
even more exacerbated in the case of shallow surveys, where the
photometric scatter, especially in the bands used for drop-out, could
undermine the completeness of the selection criteria adopted. It is
thus not surprising at all that in the past there were a number of
claims of a low space density of high-z QSOs
\citep[e.g.][]{yang16,McGreer18,kulkarni19,niida20,kim20,kimim21,FB22,onken22,jiang22,shin22,pan22,schindler23,Matsuoka23},
that are not supported by the data of this paper. A possible way out
to have a high level of the completeness, while still keeping the
efficiency of the spectroscopic surveys acceptable, is to adopt
selection methods based on machine learning and iterative removing of
low probability candidates (Calderone et al. in prep). It will be
interesting in the future to apply this method to the entire HSC
database.

\subsection{Considerations and Implications from the RUBICON survey}

A number of considerations can be drawn here from the QSO search we have
carried out on the HSC-Udeep and HSC-Wide surveys. The relevant points are:
\begin{enumerate}
\item
  The application of a magnitude threshold $magZ\le 20$ to the GAIA database
  is very efficient in rejecting stars, as shown in Fig. \ref{fig:grizudeep} and
  Fig. \ref{fig:grizwide}.
\item
  The extended dynamic ranges in the G and R bands of the HSC-Udeep and HSC-Wide
  surveys are important to select high-z QSOs, as shown in
  Fig. \ref{fig:grizudeep} and Fig. \ref{fig:grizwide}.
\item
  The spread in the G-R color is due to both redshift evolution and
  fluctuations of the IGM opacity in different lines of sight.
  Objects with blue G-R colors ($\sim 1.6$) can be at $z\ge 4.5$, as
  shown in Fig. \ref{fig:grizudeep}. In order to compute the completeness
  level and the selection function, detailed simulations have been carried out
  by including the stochasticity of the IGM absorption at these redshifts.
\item
  The G-R color selection could be biased toward QSOs with low mean
  free path or negligible escape fraction of Lyman continuum photons.
  A possible solution to this issue can be to carry out
  extensive spectroscopic confirmation of candidates that lay close to
  the border of the color selection region, outside the adopted
  criteria. The selection of objects with $G-R\le 1.6$ will allow us
  to quantify the completeness of our color criteria and the effective
  distribution of the mean free path of ionizing photons. This will be
  possible thanks to the availability of parallaxes
  and proper motions of unprecedented quality from GAIA, which allows to clean the catalog from
  contaminating stars.
\item
  Photometric redshifts tend in few cases to underestimate the
  spectroscopic redshifts of QSOs, as shown by the spectroscopically
  confirmed QSOs in Table \ref{tab:qsoz5udeep} and Table
  \ref{tab:qsoz5wide}. For this reason, the spectroscopic follow-up
  should be also extended to objects with $3.7\le z_{phot}\le 4.5$ in the future.
\item
  Photometric redshifts are computed at the mean IGM transmission of
  \citet{inoue14}. At high redshift, a large scatter of the IGM
  absorption $\tau_{IGM}$ is expected. Detailed simulations of the IGM
  variance have been carried out in order to calculate the effective
  completeness of the RUBICON survey.
\end{enumerate}

The lessons learned from the RUBICON survey is that the number of
interlopers is relatively low in the color selection region adopted
(provided that relatively deep surveys are available), and the
HSC-Wide and HSC-Udeep data are very promising to address the key
question of the space density of $\sim L^*$ AGN at $z\sim 5$. At
present, the results shown in Fig. \ref{fig:lfz5} and Table
\ref{tab:expectz5} are based only on the spectroscopically confirmed
QSO sample. In the future, the spectroscopic confirmation of all the
candidates in Table \ref{tab:qsoz5udeep} and Table \ref{tab:qsoz5wide}
will be important in order to assess the completeness of these
surveys.

%%%%%%%%%%%%%%%%%%%%%%%%%%%%%%%%%%%%%%%%%%%%%%%%%%%%%%%%%%%%%%%%%%%%%

\section{Conclusions}
\label{sec:conclusion}

In this paper we present the first results of the RUBICON survey,
aimed at constraining the space density of relatively bright QSOs
($M_{1450}\sim -27$) at $z\sim 5$, i.e. at the end of the reionization
epoch \citep{eilers18,keating20,bosman22,zhu22}.

From the ultradeep imaging in the grizY bands covering 34.7 sq. deg. in
the HSC-Udeep survey \citep{aihara22}, an almost complete sample of
three spectroscopically confirmed QSOs at $4.5<z<5.2$ and $magZ\le 20.0$
has been drawn (see Fig. \ref{fig:grizudeep} and Table
\ref{tab:qsoz5udeep}).
QSO candidates at $z\sim 5$ have been selected through the
$G-R$ vs $I-Z$ color-color criteria, as shown in Fig. \ref{fig:grizudeep}.
Bona fide stars have been excluded, based on the parallax and proper motion
information from GAIA DR3 \citep{gaiadr3}.
Two promising candidates in HSC-Udeep have $z_{phot}\sim 3.8$, and
they can also be at
$z_{spec}>4.5$, due to the tendency of our photo-z to underestimate the
spectroscopic redshifts of high-z QSOs (see Table \ref{tab:qsoz5udeep}).
The spectroscopic confirmation of these targets is currently on-going.

In order to check the reliability of our results on even larger areas,
the HSC-Udeep survey has been complemented by adding 108 sq. deg. of
wide and deep imaging in the HSC-Wide area, finding eight
spectroscopically confirmed QSOs (two are in common with the HSC-Udeep
survey) and five QSO candidates at $magZ\le 20.0$ with redshift
$z_{phot}\ge 3.5$ (see Fig. \ref{fig:grizwide} and Table \ref{tab:qsoz5wide}).
Some of these candidates are also expected to be QSOs with $z_{spec}>4.5$,
pending spectroscopic confirmation in the future.
One of the known QSOs (ID=124850) has been
confirmed by spectroscopic observations at Magellan telescope (Las
Campanas Observatory) in November 2022, triggered by the RUBICON survey.
This QSO has been independently discovered by \citet{yang23}.

The HSC-Udeep area is divided into four well separated fields (SXDS,
COSMOS, DEEP2-3, ELAIS-N1), thus minimizing cosmic variance effects
($\sim$10\%). For comparison, the Poissonian error in the same corresponding
area is $\sim$30-40\%, thus dominating the uncertainty of this
survey, due to the expected low numbers of high-z QSOs.

The luminosity function of $z\sim 5$ QSOs (Fig. \ref{fig:lfz5}) has
been computed as the inverse of the accessible cosmological volume at
$4.5\le z\le 5.2$, summed up for all the spectroscopically confirmed
sources in the HSC-Udeep survey. A completeness of 94\% has been
assumed. The same calculation has been carried out in the HSC-Wide
survey, giving comparable results in terms of QSO space density. These
$z\sim 5$ QSO luminosity functions are a factor of $\sim$2-3 times
larger than the one derived by
\citet{yang16,McGreer18,kulkarni19,niida20,kim20,kimim21,FB22,onken22,jiang22,shin22,schindler23,Harikane23,Matsuoka23}
and model 4 of \citet{pan22}. The Maximum Likelihood approach
summarized in Table \ref{tab:expectz5} confirms the results obtained
with the non-parametric luminosity function analysis. We estimated
the expected number of QSOs from the published parameterizations of
the $z\sim 5$ QSO luminosity functions. We show that all of them, with
the relevant exception of \citet{grazian22} and models 1, 2, 3 of
\citet{pan22}, predict a too low number of QSOs (and are thus
incompatible with our findings at $>2\sigma$ confidence level).

It is clear, from Fig. \ref{fig:lfz5} and from the maximum likelihood
analysis in Table \ref{tab:expectz5}, that the only viable
parameterization of the QSO luminosity function in agreement with the
present data is the one of \citet{grazian22}.
This has deep implications for the role of high-z AGN on the HI
reionization event. We assume an escape fraction of 70\% at all
redshifts and luminosities, and we integrate the QSO luminosity
function of \citet{grazian22} in the interval $-30\le M_{1450}\le -18$.
Given a mean free path of 17.4 proper Mpc at z=4.75, it turns out that
AGN are able to produce $\sim$50-100\% of the ionizing background at
the end of the reionization epoch. In particular, recent claims of a
negligible role of AGN in the HI reionization, based on earlier
luminosity function estimates \citep[e.g.][]{yang16,McGreer18,kulkarni19,niida20,kim20,kimim21,FB22,onken22,jiang22,shin22,pan22,schindler23,Harikane23,Matsuoka23}
are formally excluded at more than 2$\sigma$ level,
given our results here on the luminosity function of $z\sim 5$ QSOs.

We have provided here starting evidences in term of QSO number density
that AGN can produce a substantial amount of the required ionizing
photons to sustain the final phases of the Reionization epoch at
$z\sim 5$. In the future, the RUBICON survey can be extended both in
area, within all the HSC-Wide area ($\sim$1200 sq. deg.), and in
depth, thanks to the deep imaging already available in the HSC-Udeep
survey \citep[e.g.][]{desprez23}. If confirmed by future studies,
this will have a strong impact on the study of Reionization and on the
sources responsible for this cosmological event.

It is now time to cross the Rubicon of Reionization with QSOs:
``Alea Iacta Est'' (Caesar, 49 B.C.).

%---------------------------------------------------

\begin{acknowledgments}
We warmly thank the referee for providing the comments that allow us
to improve the quality and readability of the paper.

We acknowledge financial contribution from the grant PRIN
INAF 2019 (RIC) 1.05.01.85.09: ``New light on the Intergalactic
Medium (NewIGM)''.

A.G. and F.F. acknowledge support from PRIN MIUR project ``Black Hole winds
and the Baryon Life Cycle of Galaxies: the stone-guest at the galaxy
evolution supper'', contract 2017-PH3WAT.

AG, AB, and IS acknowledge the support of the INAF Mini Grant
2022 ``Learning Machine Learning techniques to dig up
high-z AGN in the Rubin-LSST Survey''.

The Hyper Suprime-Cam (HSC) collaboration includes the astronomical
communities of Japan and Taiwan, and Princeton University. The HSC
instrumentation and software were developed by the National
Astronomical Observatory of Japan (NAOJ), the Kavli Institute for the
Physics and Mathematics of the Universe (Kavli IPMU), the University
of Tokyo, the High Energy Accelerator Research Organization (KEK), the
Academia Sinica Institute for Astronomy and Astrophysics in Taiwan
(ASIAA), and Princeton University. Funding was contributed by the
FIRST program from the Japanese Cabinet Office, the Ministry of
Education, Culture, Sports, Science and Technology (MEXT), the Japan
Society for the Promotion of Science (JSPS), Japan Science and
Technology Agency (JST), the Toray Science Foundation, NAOJ, Kavli
IPMU, KEK, ASIAA, and Princeton University.

This paper makes use of software developed for Vera C. Rubin
Observatory. We thank the Rubin Observatory for making their code
available as free software at http://pipelines.lsst.io/.

This paper is based on data collected at the Subaru Telescope and
retrieved from the HSC data archive system, which is operated by the
Subaru Telescope and Astronomy Data Center (ADC) at NAOJ. Data
analysis was in part carried out with the cooperation of Center for
Computational Astrophysics (CfCA), NAOJ. We are honored and grateful
for the opportunity of observing the Universe from Maunakea, which has
the cultural, historical and natural significance in Hawaii.

This work has made use of data from the European Space Agency (ESA)
mission {\it Gaia} (\url{https://www.cosmos.esa.int/gaia}), processed
by the {\it Gaia} Data Processing and Analysis Consortium (DPAC,
\url{https://www.cosmos.esa.int/web/gaia/dpac/consortium}).  Funding
for the DPAC has been provided by national institutions, in particular
the institutions participating in the {\it Gaia} Multilateral Agreement.

This paper includes data gathered with the 6.5 meter Magellan
Telescopes located at Las Campanas Observatory (LCO), Chile.

\end{acknowledgments}

\facilities{Subaru:Hyper Suprime Cam, Gaia, Magellan:Baade (IMACS)}

%------------------------------------------------------------------------------

\appendix

\section{The spectral energy distributions of confirmed and candidates QSOs}

\begin{figure}
\includegraphics[width=\linewidth]{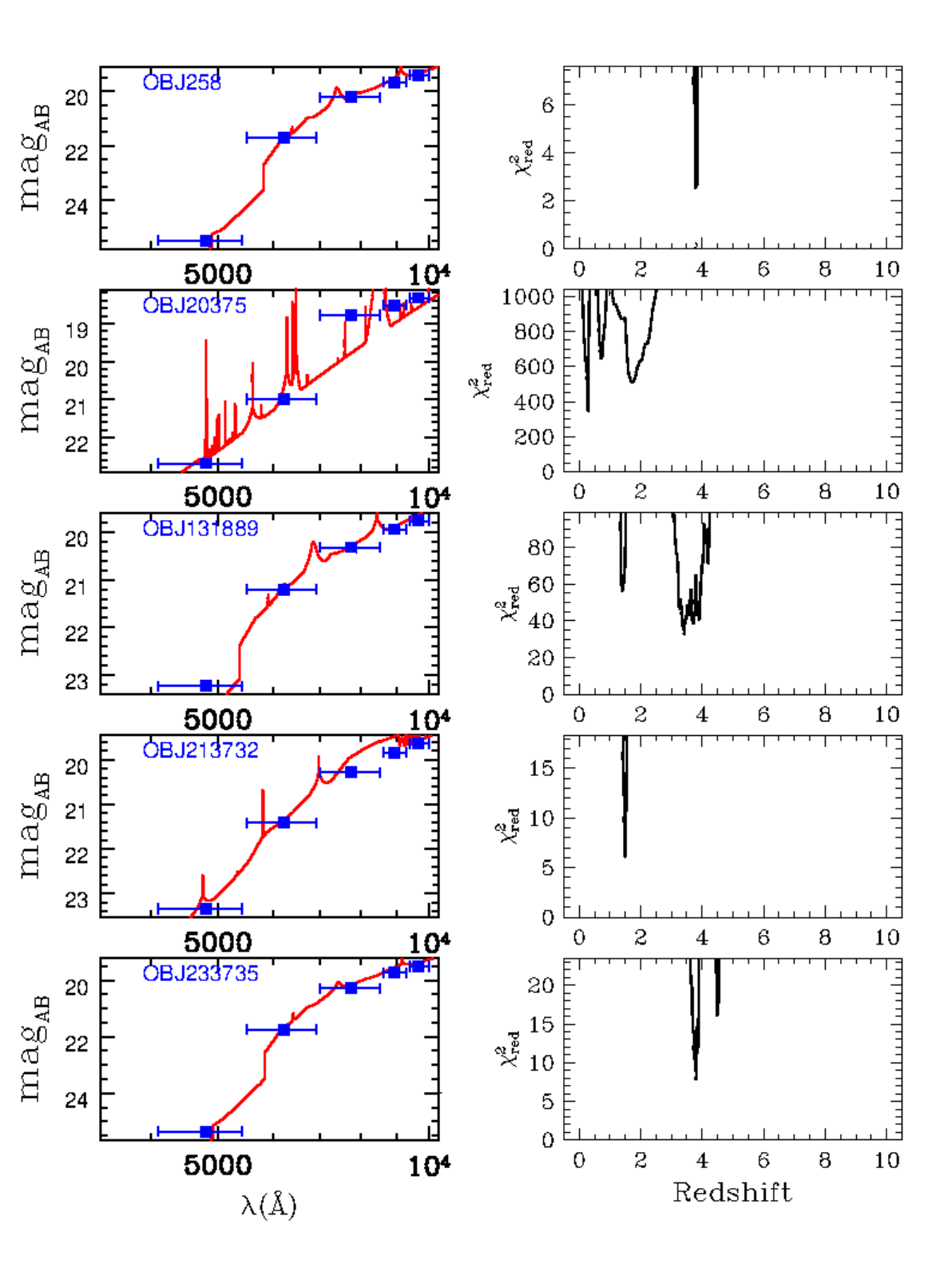}
\caption{The best-fit spectral energy distribution (left) and
the $\chi^2(z_{phot})$ at
different redshifts (right) for color selected QSO candidates in the HSC-Udeep
survey.}
\label{fig:candudeep2}
\end{figure}

\begin{figure}
\includegraphics[width=\linewidth]{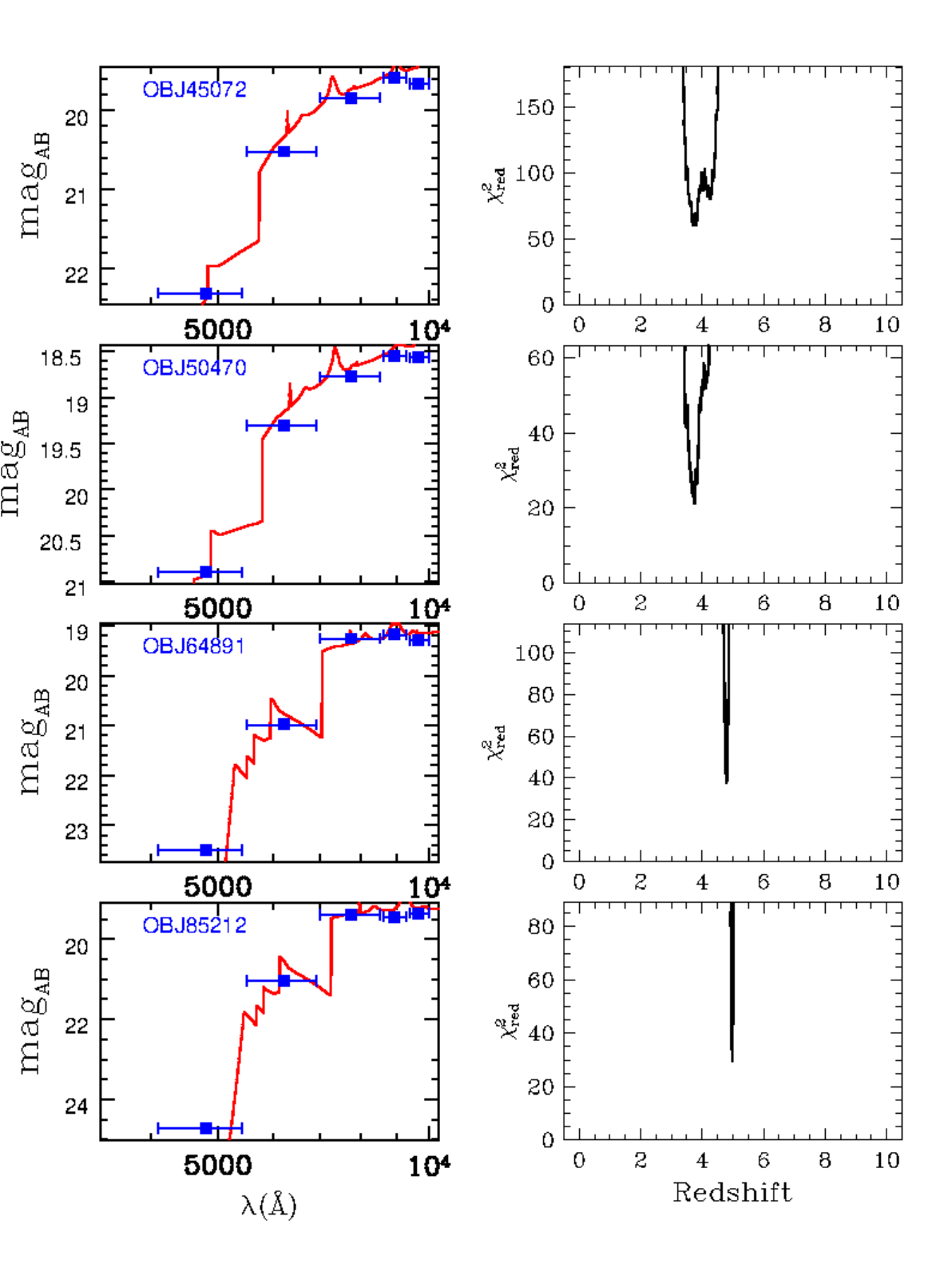}
\caption{The best-fit spectral energy distribution (left) and
the $\chi^2(z_{phot})$ at
different redshifts (right) for spectroscopically confirmed QSOs at
$z\sim 5$ in the HSC-Wide survey.}
\label{fig:candwide1}
\end{figure}

\begin{figure}
\includegraphics[width=\linewidth]{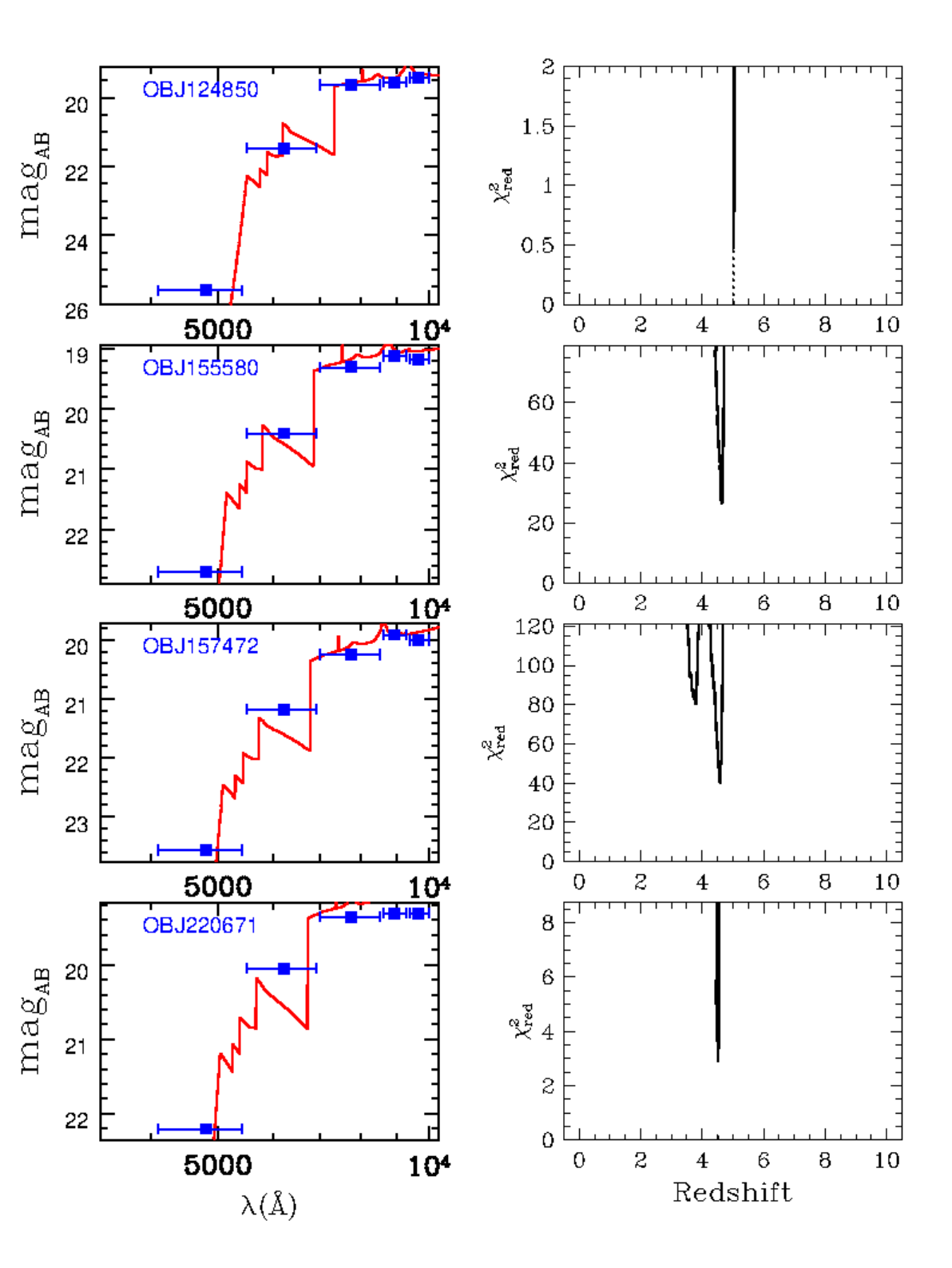}
\caption{The best-fit spectral energy distribution (left) and
the $\chi^2(z_{phot})$ at
different redshifts (right) for spectroscopically confirmed QSOs
at $z\sim 5$ in the HSC-Wide survey.}
\label{fig:candwide2}
\end{figure}

\begin{figure}
\includegraphics[width=\linewidth]{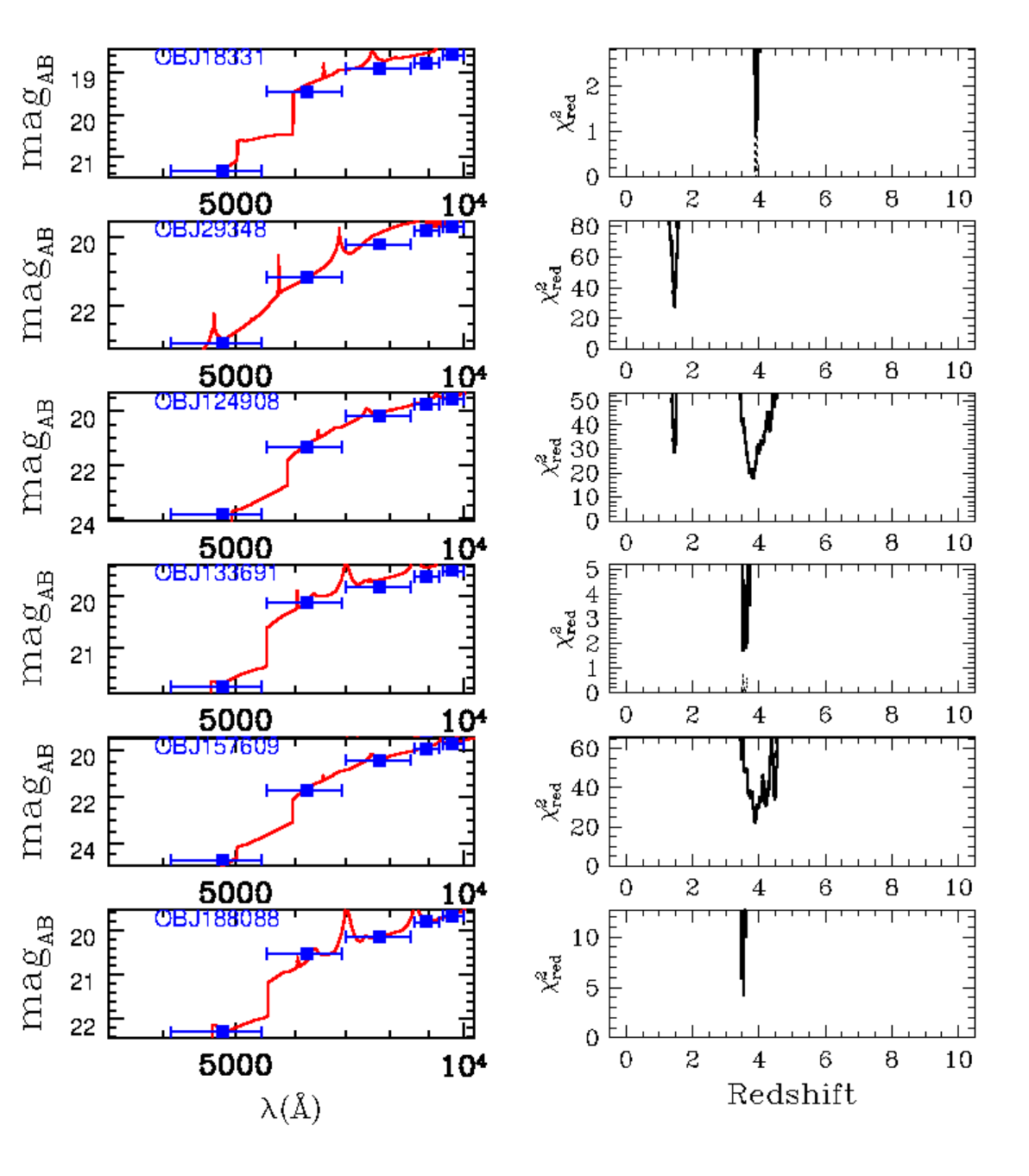}
\caption{The best-fit spectral energy distribution (left) and
the $\chi^2(z_{phot})$ at
different redshifts (right) for color selected QSO candidates in the
HSC-Wide survey.}
\label{fig:candwide3}
\end{figure}

Examples of the spectral energy distributions for QSO candidates in
the HSC-Udeep survey (Fig. \ref{fig:candudeep2}), confirmed and QSO
candidates in the HSC-Wide area (Fig. \ref{fig:candwide1},
\ref{fig:candwide2}, and \ref{fig:candwide3}).

%------------------------------------------------------------------------------

\bibliography{References}{}
\bibliographystyle{aasjournal}

\end{document}